\documentstyle[aps,multicol,pre,epsfig]{revtex}
\newcommand{\be}{\begin{equation}}
\newcommand{\ee}{\end{equation}}
\newcommand{\lb}{\ell_B}
\newcommand{\eff}{\hbox{\scriptsize eff}}
\newcommand{\sat}{\hbox{\scriptsize sat}}

\newcommand{\Zeff}{Z_{\hbox{\scriptsize eff}}}
\newcommand{\Zsat}{Z_{\hbox{\scriptsize sat}}}
\newcommand{\laeff}{\lambda_{\hbox{\scriptsize eff}}}
\newcommand{\lasat}{\lambda_{\hbox{\scriptsize sat}}}
\newcommand{\phipb}{\phi_{_{\hbox{\scriptsize PB}}}}

\newcommand{\philpb}{\phi_{_{\hbox{\scriptsize LPB}}}}
\newcommand{\kres}{\kappa_{\hbox{\scriptsize res}}}

\begin{document}
\title{Effective charge saturation in colloidal suspensions}
\author{Lyd\'eric Bocquet$^1$, Emmanuel Trizac$^2$, and Miguel Aubouy$^3$}
\address{$^{1}$ Laboratoire de Physique de l'E.N.S. de Lyon, UMR 
CNRS 5672, 46 All\'ee d'Italie, 69364 Lyon Cedex, France}
\address{$^2$Laboratoire de Physique Th{\'e}orique, UMR CNRS 8627,
B{\^a}timent 210, Universit{\'e} Paris-Sud,
91405 Orsay Cedex, France}
\address{$^{3}$ S.I.3M., D.R.F.M.C., CEA-DSM Grenoble,
17 rue des Martyrs, 38054 Grenoble Cedex 9, France}
\date{\today}
\maketitle

\begin{abstract}
Because micro-ions accumulate around highly charged colloidal particles
in electrolyte solutions, the
relevant parameter to compute their interactions is not the bare charge,
but an effective (or renormalized) quantity, whose value is sensitive to the
geometry of the colloid, the temperature or the presence of added-salt. 
This non-linear screening effect is a central
feature in the field of colloidal
suspensions or polyelectrolyte solutions. We propose a simple method to 
predict effective charges of highly charged macro-ions, that is reliable for 
mono-valent electrolytes (and counter-ions) in the colloidal limit (large
size compared to both screening length and Bjerrum length). 
Taking reference to the non linear Poisson-Boltzmann theory, the method is 
successfully tested against the geometry of the macro-ions, the possible 
confinement in a Wigner-Seitz cell, and the presence of added salt.
Moreover, our results are corroborated by 
various experimental measures reported in the literature. 
This approach provides a useful 
route to incorporate the non-linear
effects of charge renormalization within a linear theory 
for systems where electrostatic interactions
play an important role.

\end{abstract}

\pacs{PACS: 61.20.Gy, 82.70.Dd,  64.70.-p }

%\begin{multicols}{2}
%\narrowtext 

\section{Introduction}

When a solid-like object, say a colloidal particle (polyion), which carries a large
number of ionisable groups at the surface is immersed in a
polarizable medium (with a dielectric constant $\epsilon$, say water), the
ionisable groups dissociate, leaving counter-ions in the solutions and
opposite charges at the surface. The interactions between the charged
colloids, which determine the phase and structural behaviour of the
suspension, is mediated by the presence of micro-ions clouds. The complete
description of the system is thus a formidable task in general. 
However in view of the large asymmetry of size and charge between macro-
and micro- ions, one expects to be able to integrate out the micro-ions
degrees of freedom, and obtain an effective description involving macro-ions
only. In the pioneering work of Derjaguin, Landau, Verwey and Overbeek \cite
{Verwey}, micro-ions clouds are treated at the mean-field Poisson-Boltzmann
(PB) level, yielding the foundations of the prominent DLVO theory for the
stability of lyophobic colloids. An important prediction of the theory is
the effective interaction pair potential between two spherical 
colloids of radii $a$ in a solvent
which, within a linearization approximation, takes the Yukawa or
Debye-H\"{u}ckel (DH) form: 
\begin{equation}
v(r)= {\frac{Z^2 e^2}{{4 \pi \epsilon}}}\left({\frac{\exp[\kappa a]}{1+\kappa a%
}}\right)^2 {\frac{\exp (-\kappa r)}{r}},  \label{V_Yukawa}
\end{equation}
where $Z$ is the charge of the object in units of the elementary charge $e$ 
and $\kappa $ denotes the inverse
Debye screening length. The latter is defined in terms of the micro-ions
densities \{$\rho_\alpha$\} (with valences \{$z_\alpha$\}) as: $\kappa^2=4
\pi \ell_B \sum_\alpha \rho_\alpha z_\alpha^2$. The Bjerrum length $\ell_B$
is defined as $\ell _{B}=e^{2}/(4\pi \epsilon k_{B}T)$, where $\epsilon$
is the permittivity of the solvent considered as a dielectric continuum: 
$\ell _{B}=7\,$\AA\ for water at room temperature.

However, this approach becomes inadequate to describe highly charged objects
for which the electrostatic energy of a micro-ion near the colloid surface
largely exceeds $k_{B}T$, the thermal energy, because the linearization of
the PB equations is \textit{a priori} not justified. In this case however,
the electrostatic potential in exact \cite{Kjellander1,Kjellander2} 
or mean-field \cite
{Belloni,Alexander} theories still takes the Debye-H\"{u}ckel like form far
from the charged bodies, provided that the bare charge $Z$ 
is replaced by an effective or renormalized quantity $\Zeff$. 
The micro-ions which suffer a high electrostatic coupling with the colloid
accordingly accumulate in its immediate vicinity so that the 
decorated object, colloid 
\textit{plus} captive counter-ions, may be considered as a single entity
which carries an effective charge $Z_{\hbox{\scriptsize eff}}$, much lower
(in absolute value) than the structural one. Within the prominent 
mean-field PB theory \cite{Hunter} --often quite successful despite of its
limitations--, $Z$ and 
$Z_{\hbox{\scriptsize eff}}$ coincide for low values of the structural charge,
but $Z_{\hbox{\scriptsize eff}}$ eventually reaches a saturation value $%
\Zeff^{\sat}$ independent of $Z$ when the bare charge increases \cite
{Alexander,Hansen}. Arguably, the difference $Z-\Zeff$ is identified with
the amount of counter-ions ``captured'' or ``condensed'' \cite{condense}
onto the macro-ion.

A reminiscent effect has been recognized in the physics of polyelectrolytes
under the name of Manning-Oosawa condensation. Here, the object is an
infinitely long and thin rod bearing $\lambda $ charges per unit length. At
infinite dilution and in the absence of salt, it can be shown at the PB
level that the polyelectrolyte is electrostatically equivalent to a rod
carrying $\lambda_{\hbox{\scriptsize equiv}}$ charge per unit length, where
the equivalent charge density saturates to a critical value 
$\lambda_{\hbox{\scriptsize equiv}} = 1/\ell _{B}$ when $\lambda$ increases
\cite{Barrat,Katchalsky,Ramanathan1}. 
In general however, PB theory can be solved analytically in very few
geometries only and the difficulty remains to predict $Z_{\hbox{\scriptsize eff}}$
for a given colloidal system \cite{Belloni,Alexander,Hansen,Attard,Likos}. In
the absence of a general analytical framework for the computation of the
effective charge, this quantity is often considered as an adjustable
parameter to fit experimental data \cite{Gast,Stevens}. 

The aim of the present paper is to propose a method that allows to compute
effective charges comparing favorably with PB in the saturation regime,
provided the size $a$ of the charged macro-ion is much larger than 
Bjerrum length $\lb$ and screening length $\kappa^{-1}$. In the infinite
dilution limit, we define the 
effective charges from the large distance behaviour of the electrostatic
potential created by the (isolated) macro-ion \cite{rem}. 
While other definitions
have been put forward \cite{Belloni,Diehl,Kuhn,Deserno} this choice 
is relevant in view of computing a macro-ion pair potential 
at large distances, in the spirit of the DLVO approach \cite{Belloni2}.
It moreover 
avoids the ambiguity of introducing a cutoff region in space which interior
containing the colloid would exactly enclose a total charge
equal to the effective one. At leading order in curvature $(\kappa a)^{-1}$, 
our method easily provides effective charges at saturation 
close to their counterparts obtained in PB theory. In the situation of finite
colloid concentration where it is no longer obvious to extract an effective
charge from the large distance behaviour of the electrostatic potential
computed within a non-linear theory, we follow the proposition 
put forward by Alexander {\it et al.} \cite{Alexander} introducing
a Wigner-Seitz cell. In this situation, we generalize our original
method into a prescription that we successfully test in various geometries,
for different thermodynamic conditions (isolated systems or in contact
with a salt reservoir).

The paper is organized as follows. We first recall the basic framework of PB
theory in section \ref{sectionPB}. We then examine in some details the simple
case of a spherical polyion in the infinite dilution limit 
(section \ref{sectionZeff1}). This example allows us to 
devise a general method to
compute the effective charge for arbitrary colloidal systems. 
The situation of finite density of
colloids is then examined introducing Wigner-Seitz cells. The
salt free case is developed in section \ref{sectionZeff2}, while the situation of
finite ionic force is explicited in section \ref{sectionZeff3}.  
We finally confront the results obtained within our
prescription with experimental or simulation data in various geometries in
section \ref{sectionExp}. We discuss the general validity 
of our mean-field treatment relying on PB approximation
in section \ref{SectionDiscussion} and conclusions are drawn 
in section \ref{SectionConclusion}.
The preliminary results of this study have been published elsewhere \cite{Letter}.

%%%%%%%%%%%%%%%%%%%%%%%%%%%%%%%%%%%%%%%%%%%%%%%%%%%%%%%%%%%%%%%%%%%%%%%%%%%%%%%%
\section{General Framework: Poisson-Boltzmann theory}
\label{sectionPB}

Poisson-Boltzmann theory provides a mean field description of the micro-ions
clouds in the presence of the polyions, 
acting as an external potential. 
The key approximation in the approach is the neglect of 
(micro-)ionic correlations. The size of the micro-ions with density $\rho$ 
is neglected as well and the chemical potential reduces to its ideal contribution 
$\mu =k_{B}T\ln (\rho \Lambda^3 )$, where $\Lambda$ is an irrelevant length scale. 
Without loss of generality, the macro-ions are
supposed to be positively charged.

At equilibrium the \textit{electro-chemical} potential of the micro-ions is
uniform over the system. Introducing the reduced electrostatic potential 
$\phi =e\, V/k_BT$, the equilibrium condition for micro-ions thus reads at
the mean field level 
\begin{equation}
\ln (\rho^{\pm } \Lambda^3)\pm \phi =\ln (\rho _{0} \Lambda^3)
\label{ChemEqu}
\end{equation}
where $\left\{ \rho^{-},\rho^{+}\right\} $ are the density fields of the
charged micro-species (counter-ions and co-ions), which we assume for
simplicity mono-valent. The constant $\rho _{0}$ will be specified
hereafter. We restrict here to mono-valent micro-ions (both counter-ions
and salt). For higher valences, the reliability of PB deteriorates
(see section \ref{SectionDiscussion}).
The equilibrium condition, Eq. (\ref{ChemEqu}), is closed by 
Poisson's equation for the electrostatic potential 
\begin{equation}
\nabla^2 \phi =-4\pi \ell _{B}\left( \rho^{+}-\rho^{-}\right).
\label{Poisson}
\end{equation}
The gradient of equation (\ref{ChemEqu}) expresses the condition of mechanical
equilibrium for the fluid of micro-ions \cite{PRE}.
At this level, one has to separate between the no-salt and finite ionic
strength cases.

\begin{itemize}
\item  \textit{No-Salt Case} Only the released (here negative)
counter-ions are present in the system. The PB equation for the reduced
potential thus reads 
\begin{equation}
\nabla^2 \phi \,=\,\kappa ^{2}\,e^{\phi }  \label{PBnosalt}
\end{equation}
where the screening constant $\kappa $ is defined as $\kappa ^{2}=4\pi \ell
_{B}\rho _{0}$, with $\rho _{0}$ the constant introduced in Eq. (\ref
{ChemEqu}). The latter is fixed by the electroneutrality condition, which
imposes 
\begin{equation}
\int_{\cal V}d\bbox{r}\rho ^{-}(\bbox{r})=-ZeN_{c}
\label{electroneutralite}
\end{equation}
with $N_{c}$ the number of (identical) macro-ions, of charge $Ze$, contained
in the volume ${\cal V}$. The quantity
$\rho _{0}$ is a Lagrange multiplier associated with the
electroneutrality condition and has no specific physical meaning; 
it is modified by a shift of potential, which can be chosen at our convenience
to fix $\phi$ at a given point in the solution.

\item  \textit{Finite Ionic strength situation} In the finite ionic strength
case, salt is added to the solution, so that both co- and counter- ions are
present in the system. In the following we shall work in the semi-grand
ensemble, where the \{~colloids+micro-ions~\} system is put in contact with
a reservoir fixing the chemical potential of the micro-ions $\mu _{0}$. In
this case $\rho _{0}$ in Eq. (\ref{ChemEqu}) is the concentration of salt in
the reservoir (where $\phi$ is conveniently chosen to vanish), 
so that $\mu _{0}=k_{B}T\ln (\rho _{0}\Lambda^3)$. Since we are
considering mono-valent micro-ions $\rho _{0}$ coincides with the ionic
strength of the reservoir which is generally defined as 
$I_{0}=n_\alpha^{-1}\sum_{\alpha }z_{\alpha}^{2}\rho _{\alpha }^{0}$ 
for a number $n_\alpha$ of micro-ions 
species with valences $z_\alpha$ and reservoir densities $\rho_\alpha^0$.
This results in the PB equation for the reduced potential $\phi $ 
\begin{equation}
\nabla^2 \phi =\kappa ^{2}\sinh \phi   \label{PBsalt}
\end{equation}
where the screening factor $\kappa $ is now defined in terms of the
micro-ion concentration in the reservoir 
$\kappa ^{2}=8\pi \ell_{B}I_{0}\equiv \kres^2$.
\end{itemize}

In addition to these two situations, we shall also consider the case of infinite
dilution where an isolated macro-ion is immersed in an electrolyte
of given bulk salt concentration $I_0$ which thus plays the role of a reservoir.

PB equations, (\ref{PBnosalt}) or (\ref{PBsalt}), are supplemented by a set
of boundary conditions on the colloids, expressing the relationship between
the local electric field and the \textit{bare} surface charges of the
colloidal particles, $\sigma e$. This gives the boundary condition for 
$\phi$ at the surface of the colloid in the
form 
\begin{equation}
(\bbox{\nabla}\phi)\cdot \widehat {\bf n} = -4\pi \ell_B \sigma,  
\label{BCbare}
\end{equation}
where $\widehat {\bf n}$ denotes a unit vector normal to the colloid's surface.
Except in simple isotropic geometries \cite{Shkel}, the analytical solution of PB
theory is not known.

%%%%%%%%%%%%%%%%%%%%%%%%%%%%%%%%%%%%%%%%%%%%%%%%%%%%%%%%%%%%%%%%%%%%%%%%%%%%%
\section{Infinite dilution limit: asymptotic matching for the effective charge}
\label{sectionZeff1}

In this section, after recalling a few results on the planar case, we explicit our method on the
 particular example of
spheroids. We then generalize it to an arbitrary colloidal object
and consider the case of charged rods as an application. 
We work in the infinite
dilution limit, and therefore, we reject the external boundaries of the
system at infinity.

\subsection{Planar Case}

In the case of the planar geometry, the non linear PB equation can be analytically
solved. The detailed solution is given in appendix \ref{appendixA}. The important result
however is that far from the charged plane, the solution of the PB equation
reduces to that of the LPB equation :
\begin{equation}
\phi _{_{\hbox{\scriptsize PB}}}(z) \simeq \phi _{S}e^{-\kappa z}
\label{ApotapparentPlan}
\end{equation}
The  \textit{apparent} potential $\phi_S$ is equal to $\phi _{S}=4$ in the limit
of high bare charge of the plane. 

In this limit, the fixed charge boundary condition is therefore replaced on the plane 
by an effective \textit{fixed surface potential} boundary condition $\phi _{_{%
\hbox{\scriptsize LPB}}}(z=0)=\phi _{S}=4$. The effective charge
density (in the saturation -high bare charge- limit) is then computed using Gauss theorem at the surface, yielding 
\begin{equation}
\sigma _{\hbox{\scriptsize eff}}^{\hbox{\scriptsize sat}}=\frac{\kappa }{\pi l_{B}}.
\label{Achargeeffplan}
\end{equation}

\subsection{Charged spheres}

Let us now consider a highly charged isolated sphere (bare
charge $Ze$, radius $a$) immersed in a symmetric 1:1 electrolyte of bulk
ionic strength $I_{0}$. Within PB theory, the dimensionless electrostatic
potential obeys equation (\ref{PBsalt}). Suppose we know the exact
solution $\phi _{_{\hbox{\scriptsize PB}}}(r)$ (in spherical coordinates
with the origin at the center of the sphere), and the bare charge $Z$ is
large enough so that the reduced electrostatic potential at contact, 
$\phi _{_{\hbox{\scriptsize PB}}}(a)$, is (much) larger than 1. Then, we can divide
the space surrounding the polyion into two sub-regions: a non-linear region
(close to the particle's boundary) where 
$\phi _{_{\hbox{\scriptsize PB}}}(r)>1$, and a linear region  
where $\phi _{_{\hbox{\scriptsize PB}}}(r)<1$
(the potential vanishes at infinity). 
The surface delimiting these two regions is a sphere of radius $r^{*}$
such that $\phi _{_{\hbox{\scriptsize PB}}}(r^{*})\simeq 1$.

Far from colloid, the complicated non-linear effects have died out to a
substantial degree, and 
the solution also obeys the linearized Poisson-Boltzmann
(LPB) equation $\nabla^2 \phi = \kappa ^{2}\phi $, and therefore takes the
Yukawa form 
\begin{equation}
\phi _{_{\hbox{\scriptsize LPB}}}(r) \, =\,\frac{Z_{\hbox{\scriptsize eff}}}
{(1+\kappa a)} \,\ell _{B} \, \frac{e^{-\kappa \left( r-a\right) }}{r}.
\label{PhiLPBSphere}
\end{equation}
The effective charge $Z_{\hbox{\scriptsize eff}}$ is defined here without
ambiguity from the far field behaviour of 
$\phi _{_{\hbox{\scriptsize PB}}}(r)$: 
\be
\lim_{r\to \infty} \philpb(r)/\phipb(r)\,=\, 1.
\ee
In practice, $\philpb$ and $\phipb$ coincide in the linear region 
($r \gtrsim r^*$), so that $\philpb(r^*) \simeq 1$ (i.e. is a quantity
of order one). 

When $a \gg \kappa^{-1}$, the non-linear effects are confined to the immediate
vicinity of the macro-ion, with an extension $\kappa^{-1}$. We therefore have $r^*/a \simeq 1$
and as a consequence, $\philpb(r=a)\simeq \philpb(r^*) \simeq 1$. 
We thus obtain the effective boundary condition that $\philpb$ is a quantity ${\cal C}$
of order one for $r=a$; from Eq. (\ref{PhiLPBSphere}) this means that
$\Zeff^{\sat}=\,{\cal C} a\,(1+\kappa a) /\lb$. This simple argument provides
the non trivial dependence of the effective charge at saturation 
upon physico-chemical parameters; it applies in the
saturation regime of PB theory where $\Zeff=\Zeff^{\sat}$ and
assumes that the bare charge $Z$ is
high enough so that the non-linear region exists. In order to determine
the constant ${\cal C}$, we may consider the planar limit $a\to \infty$ where
the analytical solution of PB theory is known (see above and appendix \ref{appendixA}):
the surface charge density $\Zeff^{\sat}/(4\pi a^2)$ should coincide with 
that of a charged
plane $\kappa/(\pi \lb)$, Eq. (\ref{Achargeeffplan}). This imposes
that ${\cal C} = 4$ and going back to the charge: 
\begin{equation}
\Zeff^{\sat}=\,\frac{4a}{\ell _{B}}\,(1+\kappa a).
\label{eq:Zsph}
\end{equation}
In de-ionized solutions, this argument leads to the scaling $\Zeff^{\sat}
\propto a/\lb$, which has been recently tested for various latex colloids
\cite{Wette}.

The physical argument leading to (\ref{eq:Zsph}) may be rationalized as follows.
The situation of large $\kappa a$ corresponds to a low curvature limit
where the solution of Eq. (\ref{PBsalt}) may be approximated by the
solution of the planar problem in the region where curvature effects 
may be neglected : the latter corresponds to a region  $a<r< a+\delta a$, 
with $\delta a \sim a $. 
It is crucial to note that $r^*<a+\delta a $ since, as mentioned above, the extension
of the region where the non linear effects are important (defining $r^*$) has an extension
of order $\kappa^{-1}$, smaller than $\delta a\sim a$ in the limit of large
$\kappa a$. 
As a consequence, in the region $r^*<r<a+\delta a$, the solution of
the LPB equation, Eq. (\ref{PhiLPBSphere}), may be matched to the asymptotic expression of
the {\it planar} solution, given in Eq. (\ref{ApotapparentPlan}) (using $r\sim a$ and $z\simeq r-a$).
The expression (\ref{eq:Zsph}) is therefore recovered, showing again that
at the linearized level, the apparent potential is $\philpb(a) = 4$ in the saturation limit.

Eq. (\ref{eq:Zsph}) provides by construction the correct large 
$\kappa a$ behaviour of 
$\Zeff^{\sat}$, and becomes exact (compared to PB) in the planar limit.
We will show below that it remains fairly accurate down to 
$\kappa a$ of order 1. A similar expression may be found in 
\cite{Crocker,Behrens}, but the generality of the underlying method
does not seem to have been recognized.
This result is supported by the work of
Oshima \textit{et al.} \cite{Oshima} which proposes an approximation scheme
of the non-linear PB equations for spheres in infinite dilution, for large
$\kappa a$. In
particular these authors obtain an analytical approximation for the apparent
potential at the colloid surface, which reads in the saturation regime: 
\begin{equation}
\phi_{S}^{\hbox{\scriptsize Osh}} \,=\,
\,8\,\frac{1+\kappa a}{1+2\kappa a}.  
\label{eq:phi_oshima}
\end{equation}
Supplemented with expression (\ref{PhiLPBSphere}), this leads to the improved
effective charge
\begin{equation}
\Zeff^{\sat}\,=\,\frac{8a}{\ell _{B}}\,\frac{(1+\kappa a)^{2}}{1+2\kappa a}.
\label{eq:Zsph_Oshima}
\end{equation}
In the limit of large $\kappa a$ where $\phi_S \to 4$, both Eqs. 
(\ref{eq:Zsph}) and (\ref{eq:Zsph_Oshima}) have the same behaviour.

\begin{center}
\begin{figure}[h]
\epsfig{figure=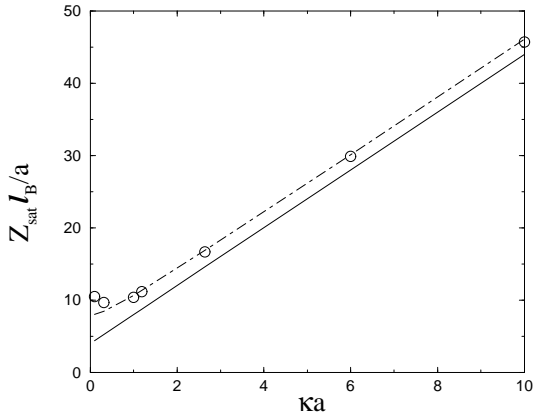,width=7cm,angle=0}
\caption{Effective charge in the saturation regime 
$\Zeff^{\sat} \lb/a$ as a function of 
$\kappa a$ for spheres in the infinite
dilution limit with added salt. The symbols (open circle) are the ``exact''
solution estimated from the large distance behaviour of the electrostatic
potential solution of the full non-linear PB equation. The continuous (resp.
dashed) line is $Z_{\hbox{\scriptsize sat}}$ found with Eq. (\ref{eq:Zsph})
[resp. Eq. (\ref{eq:Zsph_Oshima})]. }
\label{fig:sph}
\end{figure}
\end{center}

In order to test the validity of these results, we have numerically solved
the full non-linear PB equation, Eq. (\ref{PBsalt}) and computed the
effective charge from the electrostatic potential at large distances,
\textit{i.e.} the value required to match $\phi_{_{\hbox{\scriptsize LPB}}}$
to the far field $\phi_{_{\hbox{\scriptsize PB}}}$ obtained numerically. 
For each value of $\kappa a$, we make sure to consider large enough
bare charges in order to probe the saturation regime of $\Zeff$. 
Figure \ref{fig:sph} compares the numerical PB saturation 
value of the effective charge
to the prediction of our approach, Eq. (\ref{eq:Zsph}), and to that
obtained using the results of Oshima \textit{et al.}, 
Eq. (\ref{eq:Zsph_Oshima}).
We see that Eq. (\ref{eq:Zsph_Oshima}) provides an accurate estimate for 
$\Zeff^{\sat}$ as a function of $\kappa a$, for $\kappa a \gtrsim 1$. 
Working at the level of our approach only, Eq. 
(\ref{eq:Zsph}) still yields a reasonable estimate for $\Zeff^{\sat}(\kappa a)$,
specially for high values of the parameter $\kappa a$. 
In the limit of small $\kappa a$, both expressions (\ref{eq:Zsph}) and 
(\ref{eq:Zsph_Oshima}) differ notably from the PB saturation charge
which diverges, as shown by Ramanathan \cite{Ramanathan2}, as:
\be
\Zeff^{\sat} \, \sim  \, \frac{a}{\lb}\, 
\left\{ -2\ln(\kappa a) + 2 \ln[-\ln(\kappa a)] + 4 \ln 2 \right\} 
\qquad\hbox{for}\quad \kappa a \to 0.
\ee

At this point, it is instructive to briefly reconsider the work 
of Squires and Brenner \cite{Brenner} who demonstrated that the
attractive interactions between like-charged colloidal spheres
near a wall could be accounted for by a non-equilibrium hydrodynamic effect
(see also \cite{PRE}). In their analysis, they used
an ad hoc value of 0.4 for the ratio $\sigma_{\hbox{\scriptsize glass}}/
\sigma_{\hbox{\scriptsize sphere}}$ of surface charge densities
of planar and spherical polyions, in order to capture
the one-wall experiment of Larsen and Grier \cite{Larsen}. 
This was the only free parameter in their
approach. From Eqs. (\ref{eq:appsigmasat}) and (\ref{eq:Zsph_Oshima})
for the saturation values, we easily obtain 
\be
\frac{\sigma_{\hbox{\scriptsize glass}}}{\sigma_{\hbox{\scriptsize sphere}}} 
\,=\, \frac{\kappa a (1+2 \kappa a)}{2 (1+\kappa a)^2}.
\ee
In the experiment of \cite{Larsen}, we have $\kappa a \simeq 1.2$ 
[larger than 1, so that (\ref{eq:Zsph_Oshima}) is reasonably accurate],
and we obtain $\sigma_{\hbox{\scriptsize glass}}/
\sigma_{\hbox{\scriptsize sphere}} \simeq 0.42$: it is thus possible 
to justify the choice made in \cite{Brenner} assuming that both the
confining wall and the pair of colloids are charged enough to sit in the
saturation regime. In this respect, knowledge of their bare charges is 
unnecessary.

%%%%%%%%%%%%%%%%%%%%%%%%%%%%%%%%%%%%%%%%%%%%
\subsection{Arbitrary colloidal object}
\label{ssec:arbitrary}

Generalizing this analysis for an arbitrary colloidal object (of typical
size $a$), we propose the following method to estimate the effective charge
in the limit of large values of $\kappa a$:

\begin{enumerate}
\item  Solve the LPB equation for the geometry under consideration

\item  Define the saturation value, $\Zeff^{\sat}$, such that the linear
reduced potential at contact is a constant, ${\cal C}$, of order unity 
\begin{equation}
\left| \phi _{S}-\phi_{\hbox{\scriptsize bulk}}\right| ={\cal C},  
\label{ansatzDilue}
\end{equation}
where the asymptotic matching with the planar case yields ${\cal C}=4$. 

\item  If one is interested in the effective charge for arbitrary
and possibly small bare charges, a crude approximation follows from
\begin{eqnarray}
\Zeff &=&Z\,\,\,\,\,\qquad \qquad \qquad \qquad Z\ll \Zeff^{\sat} 
\nonumber \\
\Zeff &=&\Zeff^{\sat}\qquad \qquad \qquad \qquad Z\gg \Zeff^{\sat}.
\end{eqnarray}
\end{enumerate}

Our approach has several advantages. First, we do not need to solve the full
non-linear PB equations to obtain the effective charge. 
Second, the proposed method provides an analytical prediction for 
$\Zeff^{\sat}$. Third, our approach is easily adapted to other macro-ion
geometries or finite dilutions, unlike that of Ref. \cite{Oshima}
(even if these authors could find an equivalent of expression 
(\ref{eq:phi_oshima}) for cylinders, see below).

In the following, we will mainly focus on the high bare charge limit of the
colloids where the effective charge reaches a saturation plateau, 
$\Zeff^{\sat}$. In order to simplify notations, we will denote
this saturation value $Z_{\hbox{\scriptsize sat}}$.

%%%%%%%%%%%%%%%%%%%%%%%%%%%%%%%%%%%%%
\subsection{Rod-like macro-ions}

\begin{center}
\begin{figure}[h]
\epsfig{figure=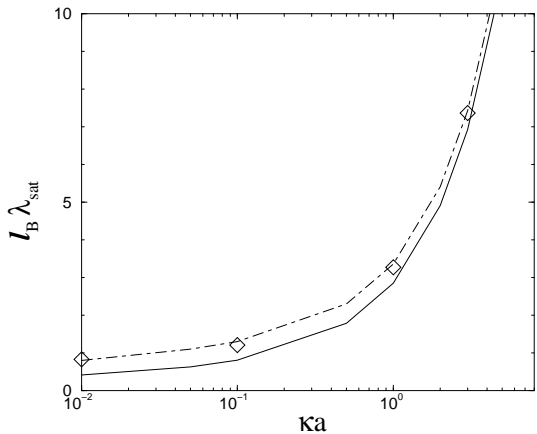,width=7cm,angle=0}
\caption{Effective line charge density, $\ell _{B}\lambda_{\hbox{\scriptsize sat}}$, 
versus $\kappa a$ (the reduced Debye-H\"uckel constant) for cylinders in the
infinite dilution limit with added salt. The symbols (open diamonds) 
are computed from the large distance behaviour of the
electrostatic potential solution of the full non-linear PB equation,
solved numerically. The continuous (resp. dashed) line is our estimate for 
$\ell _{B}\lambda_{\hbox{\scriptsize sat}}$, 
Eq. (\ref{eq:Zcyl}) [resp. the improved estimate,
Eq. (\ref{eq:Zcyl_Oshima})]. }
\label{fig:cyl}
\end{figure}
\end{center}

Now the object is an infinitely long cylinder (radius $a$, bare line
charge density $\lambda  e$). The solution of linear PB equation is (in
cylindrical coordinates where $r$ is the distance to the axis) 
\begin{equation}
\phi (r)=2\lambda \ell _{B}\,\frac{1}{\kappa a}\,\frac{K_{0}(\kappa r)}{%
K_{1}(\kappa a)},  \label{eq:potcyl}
\end{equation}
where $K_{0}$ and $K_{1}$ are (respectively) the zero and first order
modified Bessel functions of the second kind. Hence the apparent potential
is 
\begin{equation}
\phi _{S}=2\lambda \ell _{B}\,\frac{1}{\kappa a}\,\frac{K_{0}(\kappa a)}{%
K_{1}(\kappa a)}
\end{equation}
Setting $\phi _{S}={\cal C}=4$ yields our estimate for the effective
line charge density at saturation 
\begin{equation}
\lasat=\frac{2\kappa a}{\ell _{B}}\frac{K_{1}(\kappa a)}{%
K_{0}(\kappa a)}.  \label{eq:Zcyl}
\end{equation}

In the limit of large values of the bare line charge density $\lambda $, Oshima 
\textit{et al.} obtained an approximate expression for the apparent
potential in the saturation regime (based on an approximation scheme for the
PB equation, see Appendix A of Ref. \cite{Oshima}): 
\begin{equation}
\phi _{S}^{\hbox{\scriptsize Osh}}
=8\frac{K_{1}(\kappa a)}{[K_{0}(\kappa a)+K_{1}(\kappa a)]}
\label{eq:arpecOshimarod}
\end{equation}
As expected, we note that $\phi _{S}^{\hbox{\scriptsize Osh}} \to 4$ 
in the limit 
$\kappa a\gg 1$. However, from Eq. (\ref{eq:arpecOshimarod}), we deduce an
improved estimate of $\lasat$ 
\begin{equation}
\lasat\,=\,\frac{4\kappa a}{\ell _{B}}\frac{K_{1}(\kappa a)}{%
K_{0}(\kappa a)}\frac{K_{1}(\kappa a)}{[K_{0}(\kappa a)+K_{1}(\kappa a)]}.
\label{eq:Zcyl_Oshima}
\end{equation}

In Fig. \ref{fig:cyl}, we display $\ell _{B}\lasat$ 
[estimated either with Eq. (\ref{eq:Zcyl}) or Eq. (\ref{eq:Zcyl_Oshima})]
as a function of $\kappa a$, 
together with the ``exact'' value of $\ell _{B}\lasat$
found by solving the full non-linear PB equation
for high bare charges in the saturation regime. Note that the plot is in
log-linear scale, in order to emphasize the small $\kappa a$ region where
our method is not \textit{a priori} expected to work. Surprisingly,
the agreement between the numerical result of the full PB equation and 
(\ref{eq:Zcyl_Oshima}) is satisfactory down to very low values of $\kappa a$, 
$\kappa a\sim 10^{-2}$, although the two quantities have a different
asymptotic behaviour: the exact $\ell _{B}\lasat$ is finite when 
$\kappa a\to 0$ ($\ell _{B}\lasat=2/\pi$ from ref. \cite{Tracy}, see next paragraph), 
whereas both
estimate Eq. (\ref{eq:Zcyl}) and (\ref{eq:Zcyl_Oshima}) vanish, although
extremely slowly (as $-1/\log (\kappa a)$).

Importantly, $\kappa a\to 0$ is the asymptotic regime where the celebrated
Manning limiting law \cite{Ramanathan1,Fixman} happens to be exact, and the
condensation criterion holds. In this limit, above the condensation
threshold, the electrostatic potential solution of the full non-linear PB
equation is indistinguishable from that of a cylinder carrying a line
charge density $\lambda_{\hbox{\scriptsize equiv}}=1/\ell _{B}$ 
\cite{Ramanathan1}. The two
quantities $\lambda_{\hbox{\scriptsize equiv}}$ and $\lasat$ may be coined as
``effective charges'', but we maintain our initial definition of the
effective charge from the far field potential solution of the non-linear PB
equation. In this respect, $\lambda_{\hbox{\scriptsize equiv}}\neq \lasat$, 
(as already discussed in the Appendix of reference \cite{Ramanathan1}). 
This is because
one expects a remnant non-linear screening of 
$\lambda_{\hbox{\scriptsize equiv}}$, so that 
$\lasat<\lambda_{\hbox{\scriptsize equiv}}=1/\ell _{B}$. 
The limiting situation $\kappa a\to 0$ has been solved
recently within Poisson-Boltzmann theory, using exact results from the
theory of integrable systems \cite{Tracy}. The corresponding solution 
illustrates our point.
This seminal work allows to
compute explicitly the effective charge, which reads 
\begin{equation}
\lim_{\kappa a \to 0} \laeff \, = \, 
\frac{2}{\pi \ell _{B}}\,\sin \left(\frac{\pi }{2}\lambda \ell _{B}\right) .
\end{equation}
Accordingly, when $\lambda $ exceeds the Manning threshold $1/\ell _{B}$,
the effective charge saturates to a value 
\begin{equation}
\lasat\,=\,\frac{2}{\pi \ell _{B}}\,\cong \,\frac{0.6366}{\ell _{B}}%
\,<\,\lambda_{\hbox{\scriptsize equiv}}=\frac{1}{\ell _{B}}.  \label{eq:tracy}
\end{equation}
It is noteworthy that the limit $2/(\pi \ell _{B})$ (compatible with the
numerical results reported by Fixman, see for example Fig. 1 of ref. \cite
{Fixman}) is reached extremely slowly as $\kappa a$ is decreased, in
practice for $\kappa a<10^{-6}$. For example, for $\kappa a=10^{-2}$, the
numerical solution of the PB equation yields $\lasat\cong 0.81/\ell
_{B}$, hence a value 30\% larger than the asymptotic limit.

%%%%%%%%%%%%%%%%%%%%%%%%%%%%%%%%%%%%%%%%%%%%%%%%%%%%%%%%%%%%%%%%%%%%%%%%
\section{Effective charge at finite concentration. The no-salt case}
\label{sectionZeff2}

The situation of finite density of colloids does not allow to define an
effective charge from the far field of a single body potential, as done in
section \ref{sectionZeff1}. Here, we rely on the proposition put forward 
by Alexander {\it et al.} to
define an effective charge \cite{Alexander}. We recall here the main points
of this PB cell approach. First, the procedure makes use of the concept of
Wigner-Seitz (WS) cells: the influence of the other colloids is accounted
for by confining the macro-ion into a cell, with global electroneutrality
\cite{Marcus,Trizac,Deserno3,DesernoGrunberg}.
The size of the cell, $R_{WS}$ is computed from the density of colloids,
while its geometry is chosen as to mimic the spatial structure of the
colloids in the solution. Second, the ``effective'' potential solution of
the linearized PB equation is such that the linear and non-linear solutions
match up to the second derivative at the boundaries of the WS cell (hence
they match up to at least the third derivative because of
electroneutrality in ``isotropic'' --spherical or cylindrical-- cells). 
Note that in the original paper of Alexander \textit{et
al. } the procedure was introduced to obtain the effective charge from the 
\emph{numerical} solution of the non linear PB equation. But in the present
work we shall use the approach to get effective charges at the LPB level
together with our prescription. Such a route has proven successful
for mono-valent micro-ions, see e.g. \cite{Hartl,Lowen,Lobaskin}, and it has been 
shown recently that similar ideas could be employed to describe 
discrete solvent effects (again for mono-valent micro-ions 
\cite{Allahyarov}).

In this section, we generalize the analysis proposed in section \ref{ssec:arbitrary} 
to find a prescription suitable to treat the case of finite concentration of
colloids. We eventually compare our results to those obtained following ref. 
\cite{Alexander}, for planar, cylindrical and spherical geometries.

%%%%%%%%%%%%%%%%%%%%%%%%%%%%%%%%%%%%%%%%%%%%%%
\subsection{Generalized prescription and planar test case}

In the infinite dilution case, the reference potential is the bulk one
$\phi_{\hbox{\scriptsize bulk}}$. The natural
generalization of this choice for the finite concentration case consists in
replacing in Eq. (\ref{ansatzDilue}) $\phi_{\hbox{\scriptsize bulk}}$ 
by $\phi _{_{\Sigma }}$ the
reduced electrostatic potential at the boundary of the WS cell. Hence, we
propose 
\begin{equation}
\left| \phi _{S}-\phi _{_{\Sigma }}\right| ={\cal C}.  
\label{eq:presc2}
\end{equation}
If added salt was present in the suspension (see section \ref{sectionZeff3}), 
we should recover
Eq. (\ref{ansatzDilue}) from Eq. (\ref{eq:presc2})
in the infinite dilution limit where $R_{WS}$ goes to infinity.
We consequently expect the value ${\cal C}=4$ to be relevant for
the situation of finite density of colloids with added salt.
Searching for a unified description, we also test the possible validity
of the choice ${\cal C}=4$ in the no salt situation.
It is therefore instructive, as an illustration of the method and benchmark, 
to analyze the simple case of a charged plane confined in a
WS cell, without added electrolyte. As recalled in Appendix 
\ref{appendixB}, the
analytical solution of the non linear PB equation is known in such a
geometry when counter-ions are the only micro-ions present, 
which allows to check the validity of our assumptions in the
limiting case of finite concentration. Below, we compare these ``exact''
results to the predictions of our prescription.

The exact apparent potential, $\phi _{S}$ is obtained using Eq. (\ref
{pot_LPB}) at $x=0$ for the plane: $\phi _{S}=\cosh (K_{_{\hbox{\scriptsize
LPB}}}h)-1$. Our prescription imposes $\phi _{S}=4$, yielding $K_{_{%
\hbox{\scriptsize LPB}}}h=\mathrm{ArcCosh}(5)$. The effective charge is
obtained from Gauss theorem at the surface, \textit{i.e. } Eq. (\ref
{eq:Klpbnosalt}) with $\sigma $ replaced by $\sigma_{\eff}$. This leads to
the final result of our prescription $\sigma_{\sat}=\sqrt{6}\mathrm{ArcCosh}%
(5)\sigma _{c}\cong 5.6\sigma _{c}$ (where $\sigma _{c}=1/\pi l_{B}h$),
which should be compared to the exact result 
$\sigma_{\sat} \simeq 5.06\sigma _{c}$ [see Eq. (\ref{sigma_sat}) in appendix 
\ref{appendixB}].
First it is striking to note that our prescription predicts the correct
functional dependence of the effective charge in terms of the parameters of
the system. Moreover the numerical prefactor in front of $\sigma _{c}$ is
only within 10\% of the ``exact'' value obtained in Eq. (\ref{sigma_sat}),
which is quite a satisfactory agreement.

However certainly the most interesting feature which comes out from the
previous results is the fact that the apparent potential at contact, $\phi_S$, 
obtained within the analytical resolution of the PB equation, does
saturate to a constant value $\phi_S\simeq 3.66$ in the limit of very large
bare charges: this value is very close to the value we prescribe, 
$\phi_S=4$! This is a non trivial point, since the physical conditions in
the present case are very different from the isolated plane case (previous
section). 
We conclude that the analytic results
available for a confined one-dimensional electric double-layer support our
prescription. For a more refined analysis of the electrostatics of 
counter-ions between planar charged walls, going beyond
PB, we refer to the work of Netz {\it et al.} \cite{Netz}. 

In the remaining of this section, we further test our prescription against
results for spherical and cylindrical macro-ions.

%%%%%%%%%%%%%%%%%%%%%%%%%%%%
\subsection{Spheroids}

Here, the object is a charged spherical colloid (bare charge $Ze$, radius $a$) 
confined with its counter-ions in a concentric WS sphere (radius $R_{WS}$%
). The packing fraction is defined as $\eta =(a/R_{WS})^{3}$. PB
equation is again linearized around the boundary of the WS cell, yielding
Eq. (\ref{eq:lpb}) which we recall here: 
\begin{equation}
\nabla^2 \phi =K_{_{\hbox{\scriptsize LPB}}}^{2}\left( \phi +1\right).
\label{eq:lpb2}
\end{equation}
As for the planar case, the boundary conditions are $\nabla \phi (R_{WS})=0$
(electroneutrality), $\phi _{_{\hbox{\scriptsize LPB}}}(R)=0$ (because we
impose by commodity the potential to vanish at the WS cell, see Appendix B).
The solution $\phi _{_{\hbox{\scriptsize LPB}}}$ thus reads:  
\begin{equation}
\phi _{_{\hbox{\scriptsize LPB}}}(r)=-1+f_{+}{\frac{e^{K_{_{%
\hbox{\scriptsize LPB}}}r}}{r}}+f_{-}{\frac{e^{-K_{_{\hbox{\scriptsize LPB}%
}}r}}{r}}  \label{eq:phi_lpbnosalt}
\end{equation}
with 
\be
f_{\pm }=\frac{K_{_{\hbox{\scriptsize LPB}}}R_{WS}\pm 1}{2K_{_{%
\hbox{\scriptsize LPB}}}}
\,\exp (\mp K_{_{\hbox{\scriptsize LPB}}}R_{WS}).
\label{eq:fpm}
\ee 
%\begin{eqnarray}
%&\philpb(r)=&-1-\frac{1-\KLPBR_{WS}}{2\KLPBr}\,e^{-\KLPB(r-R_{WS})}\nonumber 
The charge $\Zeff$ of the colloid is obtained from the spatial derivative of $%
\phi _{_{\hbox{\scriptsize LPB}}}$ at the colloid surface: 
\be
\Zeff\,=\,\frac{a}{\ell _{B}}\,\frac{1}{K_{_{\hbox{\scriptsize LPB}}}a}
\biggl\{ (1-K_{_{\hbox{\scriptsize LPB}}}^{2}aR_{WS})
\sinh [K_{_{\hbox{\scriptsize LPB}}}(R_{WS}-a)] 
- K_{_{\hbox{\scriptsize LPB}}}(R_{WS}-a)\cosh [K_{_{\hbox{\scriptsize 
LPB}}}(Ri_{WS}-a)]\biggr\}. 
\label{eq:alexander_spheres}
\ee
At this level the screening constant $K_{_{\hbox{\scriptsize LPB}}}$ is
still unknown: it is fixed by our prescription which imposes the apparent
potential of the colloid, such that $\phi _{_{\hbox{\scriptsize LPB}}}(r=a)=%
{\cal C}=4$ with $\phi _{_{\hbox{\scriptsize LPB}}}(r)$ given in Eq. (\ref
{eq:phi_lpbnosalt}). The effective charge, $\Zsat$, is eventually computed
using Eq. (\ref{eq:alexander_spheres}). 
%The goal is to find a  $K_{_{\hbox{\scriptsize LPB}}}$ as close as 
On the other hand $K_{_{\hbox{\scriptsize PB}}}$ is
defined from the PB counter-ion density at WS boundary $\rho^-(WS)$:
\be
K_{_{\hbox{\scriptsize PB}}}^2 = 4 \pi \lb \rho^-(WS).
\label{eq:kappb}
\ee

\begin{center}
\begin{figure}[h]
\epsfig{figure=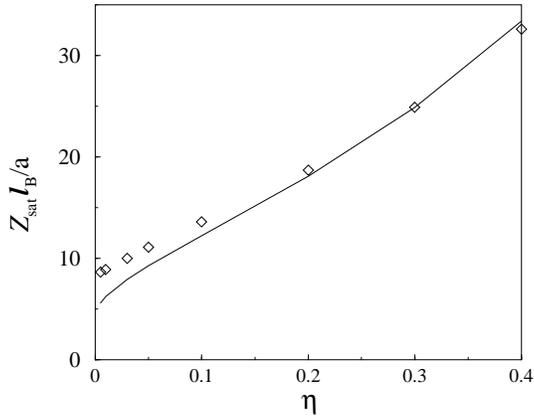,width=7cm,angle=0}
\caption{Effective charge at saturation versus  packing fraction
for spherical polyions without added salt. The symbols (open diamonds) represent the
effective charge found by solving numerically the non linear PB 
theory supplemented with Alexander's procedure \protect\cite{Alexander}. 
The continuous 
line is $Z_{\hbox{\scriptsize sat}} $ within our prescription.  }
\label{fig:sph_ws_nosalt}
\end{figure}
\end{center}

The equation for $K_{_{\hbox{\scriptsize LPB}}}$, 
$\phi _{_{\hbox{\scriptsize LPB}}}(r=a)=4$, is
solved numerically using a simple Newton procedure. Fig. \ref
{fig:sph_ws_nosalt} displays the corresponding $\Zsat$ as a function 
of $\eta$ together with the effective charge (again at saturation)
found by solving \emph{numerically} the full non linear PB equation,
together with Alexander's procedure. We recall that this procedure
defines the effective charge entering LPB equation such that the solution of
PB and LPB equations match up to the second derivative at the WS
boundary. We emphasize that once PB equation has been solved numerically,
no further numerical fitting procedure is required to match $\philpb$
to $\phipb$ and compute the effective charge: the counter-ion density at 
WS boundary $\rho^-(WS)$ is known and $K_{_{\hbox{\scriptsize PB}}}$ 
follows from  Eq. (\ref{eq:kappb}).
Replacing $K_{_{\hbox{\scriptsize LPB}}}$ with this value in 
Eq. (\ref{eq:alexander_spheres}) then gives the ``Alexander'' $\Zeff$
(a similar remark holds with added salt, see below). 
We see again that our prescription works reasonably well. 

%%%%%%%%%%%%%%%%%%%%%%%%%%%%%%%%%%%%%%
\subsection{Cylinders}
\label{ssec:rod_nosalt}

Here we apply the previous procedure to an infinite cylinder (radius $a$,
bare charge per unit length $\lambda e$) enclosed in a WS cylinder (same
axis, radius $R_{WS}$). We define the packing fraction as $\eta
=(a/R_{WS})^{2}$. This case is particularly interesting since the solution
of the PB equation in the no-salt case is known from the work of Fuoss 
\textit{et al.} and Alfrey \textit{et al.} \cite{Katchalsky}. This therefore provides another critical
test of our prescription. We note that a similar approach has been
followed by H. L\"owen \cite{Lowen}. 

The calculation follows the same lines as for the previous spherical case.
The solution of LPB equation (\ref{eq:lpb2}) in the two dimensional
case, with the usual boundary conditions and the choice 
$\philpb(R_{WS})=0$ reads: 
\be
\philpb(\rho )\,=-1+K_{_{\hbox{\scriptsize LPB}}}R_{WS}\,\biggl\{%
\hbox{I}_{1}(K_{_{\hbox{\scriptsize LPB}}}R_{WS})\,\hbox{K}_{0}(K_{_{%
\hbox{\scriptsize LPB}}}\rho )  
+\hbox{K}_{1}(K_{_{\hbox{\scriptsize LPB}}}R_{WS})\,\hbox{I}_{0}(K_{_{%
\hbox{\scriptsize LPB}}}\rho )\biggr\},  
\label{eq:phi_cyl_nosalt}
\ee
where use was made of the identity $x[I_{0}(x)K_{1}(x)+I_{1}(x)K_{0}(x)]=1$. 
From the spatial derivative of $\phi _{_{\hbox{\scriptsize LPB}}}$ at $r=a$
we deduce the effective line charge density 
$\laeff$:
\be
\laeff\,\ell _{B}=\frac{1}{2}\,K_{_{\hbox{\scriptsize LPB}%
}}^{2}aR_{WS}\biggl\{ \,\hbox{I}_{1}(K_{_{\hbox{\scriptsize LPB}}}R_{WS})\,%
\hbox{K}_{1}(K_{_{\hbox{\scriptsize LPB}}}a)
-\,\hbox{I}_{1}(K_{_{\hbox{\scriptsize LPB}}}a)\,\hbox{K}_{1}(K_{_{%
\hbox{\scriptsize LPB}}}R_{WS})\biggr\}.
\label{eq:alexander_cyl}
\ee
If the above expression is evaluated replacing $K_{_{\hbox{\scriptsize LPB}}}$
by the exact $K_{_{\hbox{\scriptsize PB}}}$ following from (\ref{eq:kappb})
once the non-linear problem has been numerically solved, we obtain
the original Alexander value (not necessarily at saturation, and without
having to implement in practice a numerical fitting procedure). 
On the other hand, as in the spherical case, the screening constant $K_{_{%
\hbox{\scriptsize
LPB}}}$ at saturation is obtained (approximately) 
by imposing the potential at the polyion's surface: 
$\philpb(a)={\cal C}=4$ in the previous equation for $\phi _{_{%
\hbox{\scriptsize LPB}}}$. Evaluation of Eq. (\ref{eq:alexander_cyl})
then gives the saturation value $\lasat$.  

\begin{center}
\begin{figure}[h]
\epsfig{figure=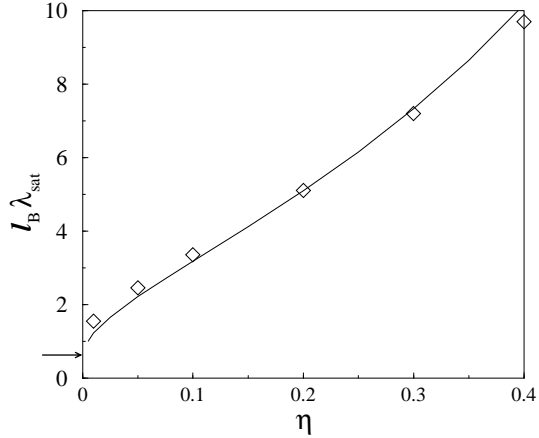,width=7cm,angle=0}
\caption{Same as Fig \protect\ref{fig:sph_ws_nosalt} for charged 
rods, except that the non linear PB results are analytical here 
\protect\cite{Katchalsky}.}
\label{fig:cyl_ws}
\end{figure}
\end{center}

\begin{center}
\begin{figure}[h]
\epsfig{figure=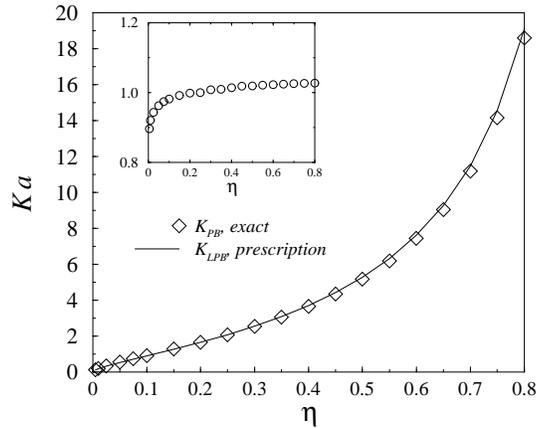,width=7cm,angle=0}
\caption{Comparison of the exact inverse screening length 
$K_{_{\hbox{\scriptsize PB}}}$ obtained from the solution of PB equation  
(diamonds) with its counterpart 
$K_{_{\hbox{\scriptsize LPB}}} $ obtained within our prescription (continuous
line), for highly charged rod-like polyions without added salt. The inset shows the
ratio $K_{_{\hbox{\scriptsize LPB}}}/K_{_{\hbox{\scriptsize PB}}}$ as
a function of packing fraction.}
\label{fig:K_cyl_ws_nosalt}
\end{figure}
\end{center}

This result is compared with the effective charge deduced by applying
Alexander's procedure to the analytical results of Fuoss \textit{et al.} and
Alfrey \textit{et al.}
in the large bare charge (saturation) limit \cite{Katchalsky}: 
the corresponding ``exact'' value for the effective charge is
chosen such that the solution of the linearized PB equation, Eq. (\ref
{eq:phi_cyl_nosalt}), matches the solution of the non linear PB equation
(Fuoss/Alfrey \textit{et al.} solution) at the WS boundary up to the second
derivative. The resulting effective charge is plotted on Fig. \ref
{fig:cyl_ws} together with the value of the effective charge obtained within
our prescription. 
In Fig. \ref{fig:K_cyl_ws_nosalt}, we compare the screening
factor $K_{_{\hbox{\scriptsize LPB}}} $ obtained within our
prescription to the ``exact'' value $K_{_{\hbox{\scriptsize PB}}} $
derived from the analytical solution of the PB equation and
Eq. (\ref{eq:kappb}) (again in
the limit of a large bare charge of the cylinder where the effective charge
saturates). The agreement between both quantities is remarkable,
even up to extremely high packing fractions (80\% on Fig. 
\ref{fig:K_cyl_ws_nosalt}).

\begin{center}
\begin{figure}[h]
\epsfig{figure=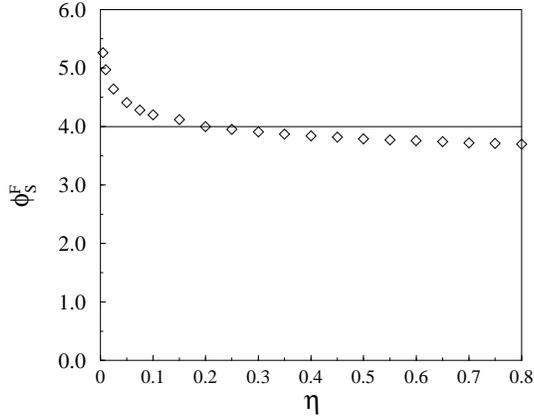,width=7cm,angle=0}
\caption{Dependence on volume fraction of the reduced linearized contact
potential, $\phi_S^{F}=\phi(a)$, with $\phi$ the LPB potential matching the
analytical solution of the PB equation, following Alexander's procedure. }
\label{fig:Phi_cyl_ws_nosalt}
\end{figure}
\end{center}

Another interesting check concerns the apparent potential at the surface of
the cylinder. Applying again Alexander's procedure to the exact 
solution of Fuoss/Alfrey \textit{et al. }, one obtains the LPB potential 
which matches the exact solution up
to its third derivative at the WS cell boundary. The value of this potential
at the surface $\phi_S^{F}$ should be compared with the value we prescribe, 
\textit{i.e. } $\phi_S=4$. The result is plotted on
Fig. \ref{fig:Phi_cyl_ws_nosalt}, showing again a good agreement 
except at low volume fraction, as expected (since as discussed
in section \ref{sectionZeff1}, our prescription is not expected to work in
the very small $\kappa a $ limit). 

Finally, we report an intriguing result: in the limit of vanishing
density it can be shown analytically that $K_{_{\hbox{\scriptsize PB}%
}}R_{WS}\to \sqrt{2}$ \cite{Katchalsky}. Using this result together with $%
K_{_{\hbox{\scriptsize LPB}}}a\to 0$, we obtain from Eq. (\ref{eq:alexander_cyl})
\begin{equation}
\lim_{\eta \to 0}\lasat=\frac{1}{\ell _{B}}\frac{\sqrt{2}}{2}\,
\hbox{I}_{1}(\sqrt{2})\simeq 0.6358\,\frac{e}{\ell _{B}}.  
\label{limit}
\end{equation}
This asymptotic value is displayed in Fig. \ref{fig:cyl_ws} with an arrow.
We observe that this limit is approached (although very slowly) as 
$\eta$ decreases. Surprisingly, the result of Eq. (\ref{limit}) is very close to
the exact expression (\ref{eq:tracy}) 
of Tracy and Widom \cite{Tracy} where the limit $\kappa a \to 0$
is taken after that of infinite dilution.
In principle, the limits of infinite dilution and of vanishing added salt
have no reason to commute. The difference between the two $\lasat$
quantities illustrates this point, with the surprise that 
the results are nevertheless very close numerically: 
\begin{eqnarray}
&&\lim_{\hbox{\scriptsize no salt}} ~~~
\lim_{\hbox{\scriptsize $\infty$ dilution}} \lasat  ~\simeq~
0.6358 \, \frac{1}{\lb}\\
&&\lim_{\hbox{\scriptsize $\infty$ dilution}} ~~~ 
\lim_{\hbox{\scriptsize no salt}} \lasat   ~\simeq~ 0.6366 \,\frac{1}{\lb}.
\end{eqnarray}

%%%%%%%%%%%%%%%%%%%%%%%%%%%%%%%%%%%%%%%%%%%%%%%%%%%%%%%%%%%%%%%%%%%%%%%%%%%%%
\section{Effective charge at finite concentration. 
The finite ionic strength case}
\label{sectionZeff3}

We now turn to the case where salt is added to the colloidal suspension.
More precisely, as already discussed in section \ref{sectionPB}, we
consider the semi grand canonical situation where the colloidal suspension
is put in contact with a reservoir of salt, through a semi-permeable membrane
(dialysis experiment). The concentration of mono-valent 
salt micro-ions in the reservoir 
$\rho _{0}$ fixes the chemical potential of the micro-ions in the
suspension. However, due to the presence of the charged macro-ions, the salt
concentration in the solution, $\rho _{s}$, differs from that in the
reservoir $\rho _{0}$: this is the so-called ``salt exclusion'' or ``Donnan
effect'' \cite{DesernoGrunberg,Dubois,PHA}.

As in the previous section the effect of finite concentration is accounted
for within PB cell theory (using a WS sphere of radius $R_{WS})$. 
Here again, we use the prescription Eq. (\ref{eq:presc2}) to predict the
effective charge of the macro-ions. For this purpose, it is convenient to
choose that the electrostatic potential $\phi $ vanishes in the reservoir
so that PB equation reads 
\begin{equation}
\nabla^2\,\phi \, =\,\kres^{2}\sinh \phi  
\label{eq:pb_res}
\end{equation}
where the screening factor $\kres$ is defined in terms of the ionic
strength of the reservoir: $\kres^{2}=8\pi \ell _{B} I_0$.

Let us now consider the linearized (``LPB'') version of this equation. We
again linearize around the value of the
potential at the boundary of the WS cell, $\phi_{_\Sigma}=\phi(R_{WS})$,
often referred to as the ``Donnan potential''
\begin{equation}
\nabla^2 \delta\phi \,=\,K_{_{\hbox{\scriptsize LPB}}}^{2}( \delta
\phi+\gamma_{0})  
\label{eq:lpbsalt}
\end{equation}
where we introduced $\delta\phi=\phi-\phi_{_\Sigma}$ and: 
\begin{eqnarray}
&&K_{_{\hbox{\scriptsize LPB}}}^{2}\,=\,\kres^{2}\cosh [\phi_{_\Sigma}] \\
&&\gamma _{0}\,= \tanh [\phi_{_\Sigma} ]\,=\sqrt{1-\left( 
\frac{\kres}{ K_{_{\hbox{\scriptsize LPB}}}}\right) ^{4}}  \label{eq:gamma0} .  
\label{eq:Kcarresel}
\end{eqnarray}
The second order differential equation (\ref{eq:lpbsalt}) is solved invoking 
the two self consistent boundary conditions
\be
\delta \phi = 0 \qquad \hbox{and} \qquad 
\frac{\partial \delta \phi}{\partial r} = 0 \qquad \hbox{for} \quad
r=R_{WS},
\ee
so that $\delta \phi$ is known as a function of distance and depends
parametrically on $K_{_{\hbox{\scriptsize LPB}}}$.
We emphasize that $K_{_{\hbox{\scriptsize LPB}}}$ is still unknown at this
point. It is computed as in section \ref{sectionZeff2}
from our prescription on the reduced potential
\begin{equation}
\delta\phi_S=\phi_S-\phi_{_\Sigma}=4.
\label{presc3}
\end{equation}

%%%%%%%%%%%%%%%%%%%%%%%%%%%%%
\subsection{Spheroids}
\label{ssec:sphere_I}

\begin{center}
\begin{figure}[h]
\epsfig{figure=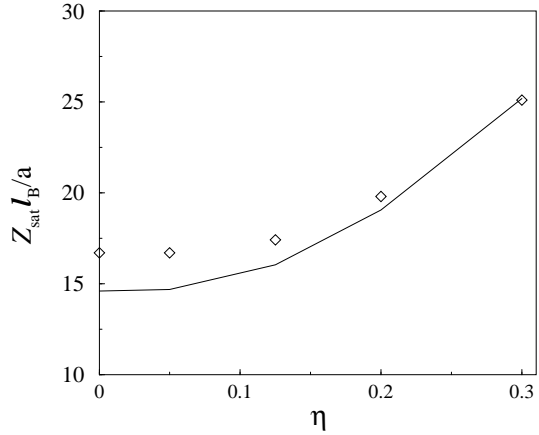,width=7cm,angle=0}
\caption{Effective charge (at saturation) of spherical colloids (radius $a$) 
as a function
of volume fraction $\eta $ for $\kres a=2.6$. The continuous line is the
effective charge (at saturation) computed using the prescription, while the 
symbols are the
results of the non-linear PB cell theory, following Ref. 
{\protect\cite{Alexander}}.  }
\label{fig:sph_ws_salt}
\end{figure}
\end{center}

With the same notations as above, the appropriate
solution of LPB equation (\ref{eq:lpbsalt}) is 
\begin{equation}
\delta \phi_{_{\hbox{\scriptsize LPB}}} (r)
=\gamma_0 \left[-1+f_{+}{\frac{e^{K_{_{\hbox{\scriptsize LPB}%
}} r}}{r}}+f_{-}{\frac{e^{-K_{_{\hbox{\scriptsize LPB}}} r}}{r}} \right]
\label{eq:phi_lpbsalt}
\end{equation}
where the functions $f_{\pm}$ are defined in Eq. (\ref{eq:fpm}).
Note that expression (\ref{eq:phi_lpbnosalt}) is recovered by taking the
formal limit $\kres =0$ in the previous equation.

Our prescription allows to compute $K_{_{\hbox{\scriptsize LPB}}}$ at saturation, 
such that $\delta \Phi (a)=4$, without any reference to the solution of the non
linear PB problem. This equation is solved numerically using a 
Newton procedure. Once $K_{_{\hbox{\scriptsize LPB}}}$ is known,
the effective charge follows from the gradient of $\delta \phi (r)$ 
in Eq. (\ref{eq:phi_lpbsalt}) taken at $r=a$ (it may also be computed
by integrating the corresponding LPB charge density over the
volume accessible to the micro-ions, i.e. $a \leq r \leq R_{WS}$)
\be
\Zsat=\gamma _{0}\frac{a}{\ell _{B}}\,\frac{1}{K_{_{%
\hbox{\scriptsize
LPB}}}a}\biggl\{ (K_{_{\hbox{\scriptsize LPB}}}^{2}aR_{WS}-1)\sinh [K_{_{%
\hbox{\scriptsize LPB}}}(R_{WS}-a)] 
+K_{_{\hbox{\scriptsize LPB}}}(Ri_{WS}-a)\cosh [K_{_{\hbox{\scriptsize LPB}%
}}(Ri_{WS}-a)]\biggr\}   
\label{eq:Zeff_ws_sphere}
\ee
with $\gamma _{0}=\sqrt{1-(\kres /K_{_{\hbox{\scriptsize LPB}}})^{4}}$. 
Again, our prescription $\delta\phi_S=\phi_S-\phi_{_\Sigma}=4$ provides
a value for $K_{_{\hbox{\scriptsize LPB}}}$ which is an approximation
for the exact $K_{_{\hbox{\scriptsize PB}}}$ at saturation, 
related to microions densities at the WS boundary through the expected
Debye-like form
\be
K_{_{\hbox{\scriptsize PB}}}^2 \,=\, 4 \pi \lb \left[\rho^+(WS) + \rho^-(WS)
\right].
\label{eq:KTr}
\ee
If Eq. (\ref{eq:Zeff_ws_sphere}) is evaluated with the exact
$K_{_{\hbox{\scriptsize PB}}}$, Alexander's original effective charge
follows (hence without having to implement any numerical fitting procedure).
We also emphasize that as in the previous sections, the right hand side
of Eq. (\ref{eq:Zeff_ws_sphere}) provides the effective $\Zeff$ 
{\it \`a la} Alexander (i.e not
necessarily at saturation), once evaluated with the correct 
$K_{_{\hbox{\scriptsize PB}}}$ (deduced from the numerical solution of the 
non-linear problem). 

\begin{center}
\begin{figure}[h]
\epsfig{figure=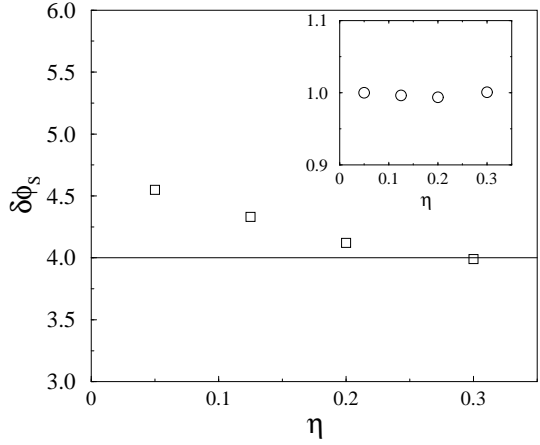,width=7cm,angle=0}
\caption{Dependence on volume fraction of the ``exact'' reduced linearized contact
potential, $\delta \phi _{S}=\phi (a)-\phi (R_{WS})$, with $\phi $ the LPB
potential matched to the numerical PB solution according to Alexander's
procedure. Our prescription assumes a constant value, $\delta \phi _{S}=4$.
Inset: ratio $K_{_{\hbox{\scriptsize LPB}}}/K_{_{\hbox{\scriptsize PB}}}$
versus packing fraction. The ratio is seen to be very close to unity over the
explored packing fraction window.}
\label{fig:sph_ws_salt_phi}
\end{figure}
\end{center}

The results for the effective charge at saturation $\Zsat$ 
as a function of volume fraction are displayed on Fig. \ref
{fig:sph_ws_salt} for $\kres a=2.6$. As in the previous sections, we
compare this result with its Alexander's counterpart at saturation. 
Our predictions are seen to be compatible with those obtained
in the PB cell model.

\begin{center}
\begin{figure}[h]
\epsfig{figure=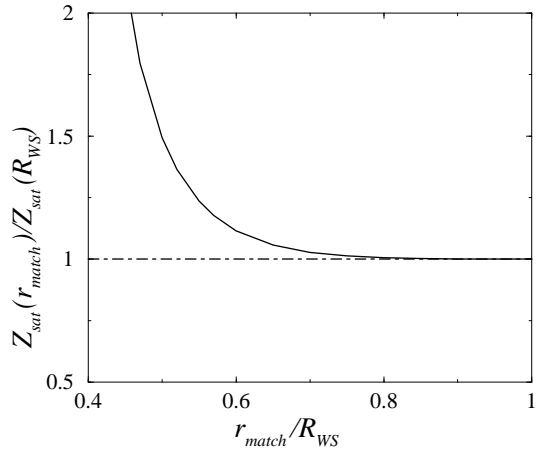,width=7cm,angle=0}
\caption{Influence of the point $r_{\hbox{\scriptsize match}}$ chosen
to match the analytical LPB and numerical PB solutions on the effective charge
in the saturation regime. The situation is that of a spherical polyion in a spherical
WS cell, at packing fraction $\eta=0.05$ and $\kres \,a=2.6$. }
\label{fig:testAlex}
\end{figure}
\end{center}

From the numerical solution of the PB equation, it is possible
to extract the apparent surface potential, $\delta \phi _{S}=\phi (r=a)-\phi
(R_{WS})$ (in the latter expression $\phi $ is defined as the
solution of the LPB equation matching the full -numerical- PB equation up to
second derivative at the WS cell boundary). By construction, 
this potential may be obtained inserting the numerically obtained
$K_{_{\hbox{\scriptsize PB}}} \equiv \kres \cosh^{1/2}(\phi_{_\Sigma})$
into (\ref{eq:phi_lpbsalt}). 
This apparent potential should
be compared against our prescription $\delta \phi =4$. The corresponding
result is shown on Fig. \ref{fig:sph_ws_salt_phi}. We observe that $\delta
\phi _{S}$ indeed saturates to a value close to $4$. The inset shows $%
K_{_{\hbox{\scriptsize LPB}}}/K_{_{\hbox{\scriptsize PB}}}$ versus $\eta $,
where $K_{_{\hbox{\scriptsize PB}}}$ is the ``exact'' screening length for
the LPB equation at saturation, obtained numerically;
$K_{_{\hbox{\scriptsize LPB}}}$ is the same quantity estimated from our
prescription. We observe that although for small packing fractions $\delta
\phi _{S}$ slightly departs from our approximation ${\cal C}=4$, the
estimated $K_{_{\hbox{\scriptsize LPB}}}$ is still remarkably close to the exact 
one.

Independently of our prescription, we finally test the relevance 
of Alexander's procedure \cite{Alexander} in the following way. 
$\Zsat$ has been obtained
above from the matching of a generic LPB potential to the numerical PB
one at $r_{\hbox{\scriptsize match}} = R_{WS}$. It is also possible to 
implement the matching at a different location inside the cell,
and we denote $\Zsat(r_{\hbox{\scriptsize match}})$ the associated
effective charge, at saturation. This quantity, normalized by the
``usual'' one $\Zsat(R_{WS})$ is displayed in 
Fig. \ref{fig:testAlex}. For the packing fraction of 5\% considered,
$r_{\hbox{\scriptsize match}} / R_{WS}$ is bounded below by
$a/R_{WS} \simeq 0.37$, and $\Zsat(R_{WS}) \simeq 16.7 a/\lb$, see
Fig. \ref{fig:sph_ws_salt}. We observe that $\Zsat$ is relatively insensitive
to $r_{\hbox{\scriptsize match}}$ for 
$0.6 R_{WS} \leq r_{\hbox{\scriptsize match}} \leq R_{WS}$.

%%%%%%%%%%%%%%%%%%%%%%%%%%%%%%%%%%%%%%%%%%%%%%
\subsection{Rod-like polyions}
\label{ssec:rod_salt}

Using the same notations as in section \ref{sectionZeff2}, the appropriate
solution of the LPB equation in cylindrical geometry reads 
\be
\delta \phi_{_{\hbox{\scriptsize LPB}}}(\rho )\,=\gamma _{0}\Biggl\{K_{_{%
\hbox{\scriptsize LPB}}}R_{WS}\,\biggl[\hbox{I}_{1}(K_{_{\hbox{\scriptsize
LPB}}}R_{WS})\,\hbox{K}_{0}(K_{_{\hbox{\scriptsize LPB}}}\rho )
+\hbox{K}_{1}(K_{_{\hbox{\scriptsize LPB}}}R_{WS})\,\hbox{I}_{0}(K_{_{%
\hbox{\scriptsize LPB}}}\rho )\biggr]-1\Biggr\}.  
\label{eq:phi_cyl_salt}
\ee
with $\gamma _{0}=\sqrt{1-(\kres/K_{_{\hbox{\scriptsize LPB}}})^{4}}$. 
Again, $K_{_{\hbox{\scriptsize LPB}}}$ is obtained as the solution of the
equation $\delta \philpb(\rho =a)=4$. Once this equation is solved, the
saturation value of the effective charge, $\lambda_{\hbox{\scriptsize eff}}$%
, follows from the spatial derivative of the potential $\delta \phi _{_{%
\hbox{\scriptsize LPB}}}$ at the rod surface: 
\be
\laeff\,=\frac{1}{2\ell _{B}}\,K_{_{\hbox{\scriptsize LPB}%
}}^{2}aR_{WS}\,\gamma _{0}\Biggl\{ \,\hbox{I}_{1}(K_{_{\hbox{\scriptsize LPB}%
}}Ri_{WS})\,\hbox{K}_{1}(K_{_{\hbox{\scriptsize LPB}}}a) 
-\,\hbox{I}_{1}(K_{_{\hbox{\scriptsize LPB}}}a)\,\hbox{K}_{1}(K_{_{%
\hbox{\scriptsize LPB}}}Ri_{WS})\Biggr\}.
\label{eq:Zcylsalt}
\ee

As in the previous sections, Eq. (\ref{eq:Zcylsalt}) gives analytically 
the relation between the effective charge {\it \`a la Alexander et al.} 
and micro-ions densities at the WS boundary. As such, it applies for
any value of the bare charge $\lambda$, and in particular, for $\lambda \to 0$,
$K_{_{\hbox{\scriptsize LPB}}}$ is such that $\laeff/\lambda \to 1$.
A similar remark applies for Eqs. (\ref{eq:alexander_spheres}), 
(\ref{eq:alexander_cyl}) and (\ref{eq:Zeff_ws_sphere}). If we choose for
$K_{_{\hbox{\scriptsize LPB}}}$ the ``exact'' $K_{_{\hbox{\scriptsize PB}}}$
value, we recover the ``exact'' cell model (Alexander) effective charges.
However, the quantity $K_{_{\hbox{\scriptsize LPB}}}$ solution of
$\delta \philpb(\rho =a)=4$ is supposedly the inverse screening length at 
saturation and therefore provides an approximation of $\lasat$ once inserted
into (\ref{eq:Zcylsalt}).

\begin{center}
\begin{figure}[h]
\epsfig{figure=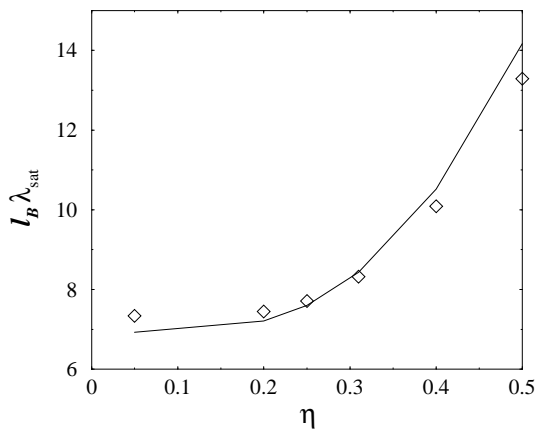,width=7cm,angle=0}
\caption{Effective charge $\ell _{B}\lambda_{\hbox{\scriptsize sat}} $ as a
function of packing fraction for cylinders with added salt 
($\kres \,a=3$). The symbols represent the effective charge
at saturation within the usual PB cell approach, while the
continuous line follows from our prescription. }
\label{fig:cyl_ws_I}
\end{figure}
\end{center}

The corresponding results for 
$\lasat$ as a function of volume fraction are displayed in
Fig. \ref{fig:cyl_ws_I} for $\kres a=3$. As in the previous sections,
we compare this result with its counterpart obtained from the numerical
solution of PB theory together with Alexander's procedure for the
effective charge in the saturation limit. The agreement with the numerical
results of the full non linear PB equation is seen to be satisfactory.

%%%%%%%%%%%%%%%%%%%%%%%%%%%%%%%%%%%%%%%%%%%%%%%%%%%%%%%%%%%%%%%%%%%%%%%%%%%
\section{Confrontation to experimental and numerical results}
\label{sectionExp} 

In the previous sections, we have tested our results for the effective charges
against the numerical solutions of PB theory. However, the effective charge
is a difficult quantity to measure directly in an experiment
(see however the work reported in \cite{Wette} confirming the scaling
$\Zsat \propto a/\lb$ for low ionic strength suspensions of spherical latex
colloids).
In order to assess the experimental relevance of the above ideas, 
we now turn to the computation of osmotic properties for  
spherical and rod-like macro-ions, easily accessible both experimentally
and within our approach. In the case of spherical colloids, we start by
considering the phase behaviour of the suspension as a function of
added salt.

%%%%%%%%%%%%%%%%%%%%%%%%%%%%%%%%%%%%%%%%%%%%%%%%
\subsection{Crystallization of charged spheres}

The phase diagram of charged spherical colloids has been widely explored
experimentally, in particular  
by Monovoukas and Gast \cite{Gast}. In this work, the
macro-ions were charged polystyrene spheres, with radius $a\simeq 660\,$\AA . The
authors moreover compared  their experimental phase diagram to
that computed for particles interacting through a Yukawa potential 
(\ref {V_Yukawa}) (the Yukawa phase diagram has indeed 
been investigated extensively by numerical simulations 
\cite{Robbins,Meijer,Bitzer}). However such a comparison experiment/theory
requires an ad-hoc choice for the
effective charge $\Zeff$ [prefactor of Eq. (\ref{V_Yukawa})]. The authors
found that a reasonable agreement with the numerical results was obtained
for a specific choice of the effective charge, $\Zsat=880$ (although they
reported conductimetry experiments indicating a macro-ion charge
around $1200$).

We focus in the following on the melting line of the phase diagram obtained 
in \cite{Gast}. We use here our predictions for the effective 
charges at saturation: we do not need to know the bare charge of the
polystyrene spheres, as this quantity is presumably much larger than 
the corresponding saturation plateau of $\Zeff$, which means,
within the PB picture, that $\Zeff \simeq \Zsat$.
Once $\Zsat$ (and the corresponding screening
constant $K_{_{\hbox{\scriptsize LPB}}}$, see previous section) is known for
a given density and ionic strength, it is possible to insert it into the
computed generic phase diagram of Yukawa systems  \cite{Robbins} 
to obtain the corresponding stable phase. We extract
in particular the melting line from these numerical results: we prefer to
use these numerical results for the phase diagram (instead of performing a
full theoretical analysis) since our main focus remains to check the
relevance of our predictions for the effective charges. This requires a
``reliable'' description for the melting line, which numerical simulations 
provide once the potential is given.

\begin{center}  
\begin{figure}[h]
\epsfig{file=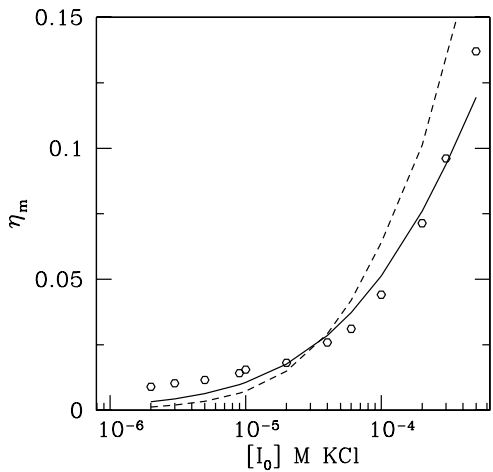,width=7cm}
\caption{Liquid-solid transition of charged polystyrene colloids: volume
fraction for melting $\eta_m$ as a function of salt ionic strength $I_0$.
Dots are experimental points for the melting line extracted from Ref.{%
\protect\cite{Gast}}. The solid line is the theoretical prediction for the
melting transition using our prescription for effective charges (see text)
while the dashed line corresponds to an ad hoc fixed
effective charge $Z_{\hbox{\scriptsize eff}}=880$, as proposed in
{\protect\cite{Gast}}. }
\label{spheres}
\end{figure}
\end{center}

We emphasize that at this level, the only parameter entering our description
is the diameter of the beads, which is measured independently. 
Accordingly, there is \textit{no adjustable parameter} in our equations
and the resulting phase diagram is strongly constrained. The results for the
melting line using our prescription for the effective charge are confronted
to the experimental data in Fig. \ref{spheres}. We also plot the result for
the melting line for an ad hoc constant effective charge, $\Zsat=880$, as was
proposed in Ref. \cite{Gast}. The observed agreement
supports the pertinence of our prescription for 
$\Zsat$ which reproduces the experimental phase diagram.
In our case, the effective charge does vary between 500 and 2000
along the melting line, depending on ionic strength and density. This could
explain that the conductimetry measurements performed independently by
Monovoukas and Gast (although at an unspecified ionic strength) yield
another value for the effective charge of the spheres ($Z\sim 1200$ as
quoted above).

%%%%%%%%%%%%%%%%%%%%%%%%%%%%%%%%%%%%%%%%%%%%%%%%%%
\subsection{Osmotic pressure of a suspension of spherical colloids}

\begin{center}
\begin{figure}[h]
\epsfig{figure=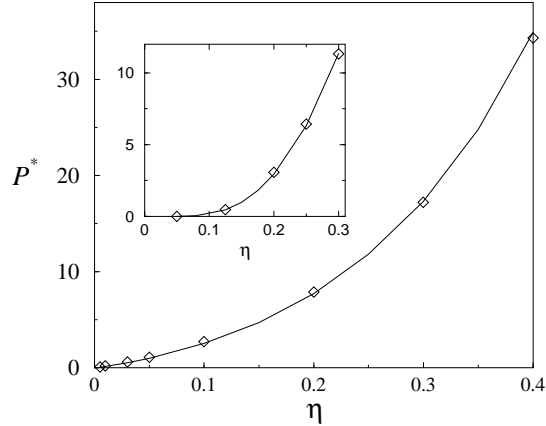,width=7cm,angle=0}
\caption{Reduced osmotic pressure 
$P^* = 4 \pi \lb a^2 \Pi _{\hbox{\scriptsize osm}} /kT$  
versus volume fraction for spherical
polyions in the salt free case where $I_0$ (and thus $\kres$) 
vanishes. The symbols are the PB values 
obtained from the resolution of the 
non-linear problem, and the line follows from our prescription. 
The inset shows the same quantities in presence of an electrolyte
($\kres a  = 2.6$).}
\label{fig:P_sph}
\end{figure}
\end{center}

\begin{center}
\begin{figure}[h]
\epsfig{figure=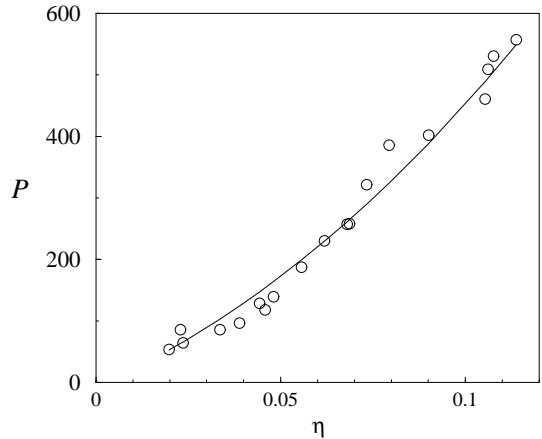,width=7cm,angle=0}
\caption{Osmotic pressure (in Pa) as a function of volume fraction.
The symbols are the experimental measures of \protect\cite{Reus}
for aqueous suspensions of bromopolystyrene particles (with radius
$a=51\,$nm). The continuous
curve corresponds to our prescription assuming a salt concentration
of $10^{-6}\,$M in the reservoir. }
\label{fig:P_Reus}
\end{figure}
\end{center}

In the PB cell model, the osmotic pressure in the solution
is related to the densities of micro-ions at the WS cell
boundary \cite{Hansen,DesernoGrunberg,Trizacbis,HansenPL}: 
\begin{equation}
\Pi _{\hbox{\scriptsize osm}}=
k_{B}T\left[\rho^{+}(R_{WS})+\rho^{-}(R_{WS})-2\rho _{0}\right],
\label{Posm1}
\end{equation}
where we have subtracted the ionic contribution from the reservoir
(of salt density $\rho_0=I_0$). 
This is because the electric field vanishes at the WS cell and there
is no contribution from the electrostatic pressure at $r=R_{WS}$. Using 
Eq. (\ref{eq:KTr}), Eq. (\ref{Posm1}) may be recast into
\begin{equation}
\Pi _{\hbox{\scriptsize osm}}={\frac{k_{B}T}{4\pi \ell _{B}}}
[K_{_{\hbox{\scriptsize PB}}}^{2}(\rho ,I_{0})-\kres^{2}],  
\label{Posm}
\end{equation}
where $\kres$ is the screening constant defined in terms of the 
ionic strength in the reservoir. Our prescription is supposed to
give an excellent approximate of the non-linear 
$K_{_{\hbox{\scriptsize PB}}}$ through $K_{_{\hbox{\scriptsize LPB}}}$,
and readily allows an estimation of the osmotic pressure.
Figure \ref{fig:P_sph} shows the accuracy of the estimate, with or without
added electrolyte.

The comparison of our predictions {\it at saturation} 
to the experimental 
results reported by Reus {\it et al.} \cite{Reus} is also satisfactory,
see Fig. \ref{fig:P_Reus}.
It was already pointed out in \cite{Reus} that PB cell theory 
reproduced well the experimental data. The agreement observed in 
Fig. \ref{fig:P_Reus} however illustrates the relevance of the PB saturation 
picture --well captured by our approach-- at high polyion/micro-ion electrostatic 
coupling (see the discussion in section \ref{SectionDiscussion}).

%%%%%%%%%%%%%%%%%%%%%%%%%%%%%%%%%%%%%%%%%%%%%%%%%%
\subsection{Osmotic properties of rod-like polyions}

\begin{center}
\begin{figure}[h]
\epsfig{figure=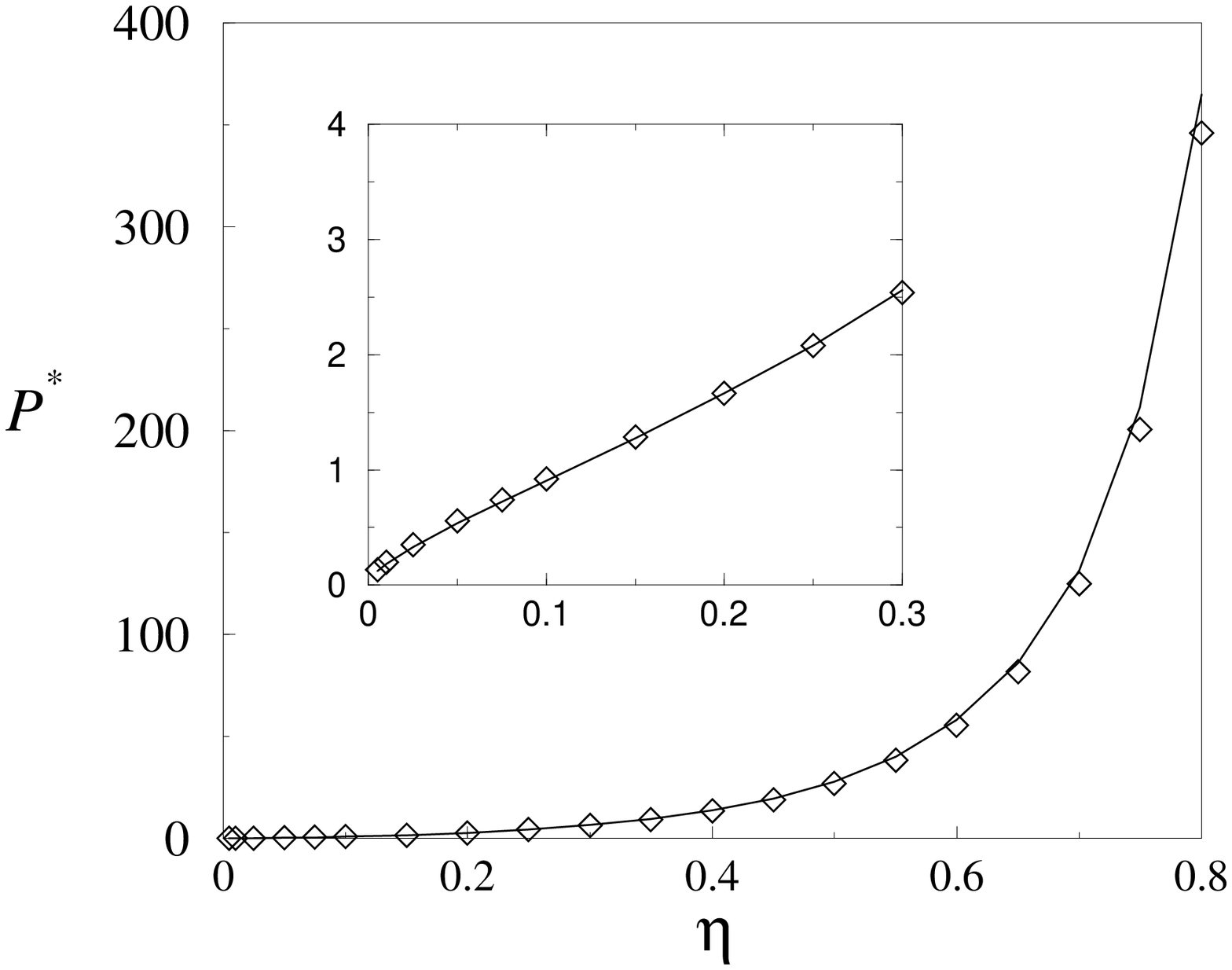,width=7cm,angle=0}
\epsfig{figure=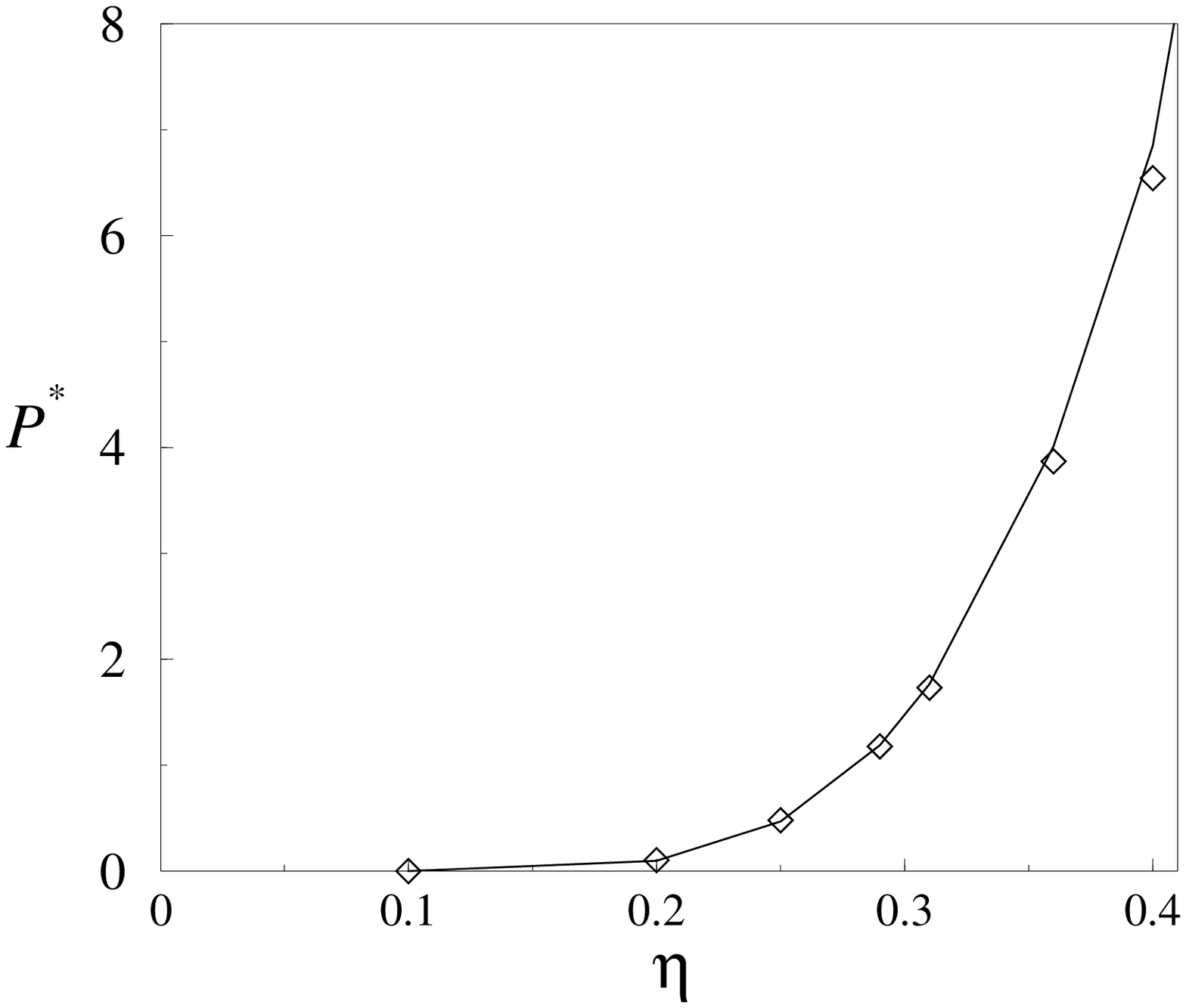,width=6.5cm,angle=0}
\caption{Same as Fig. \protect\ref{fig:P_sph} for cylindrical polyions.
Top: salt free suspensions (the inset is a zoom in the small 
packing fraction region). Bottom: situation with added salt ($\kappa a = 3$). }
\label{fig:P_cyl}
\end{figure}
\end{center}

Expression (\ref{Posm}) is also valid in cylindrical geometry, and we 
show in Fig. \ref{fig:P_cyl} the comparison prescription versus non-linear 
PB osmotic pressure. We draw a similar conclusion as for spherical
polyions concerning the accuracy of our approximation.

\begin{center}
\begin{figure}[h]
\epsfig{file=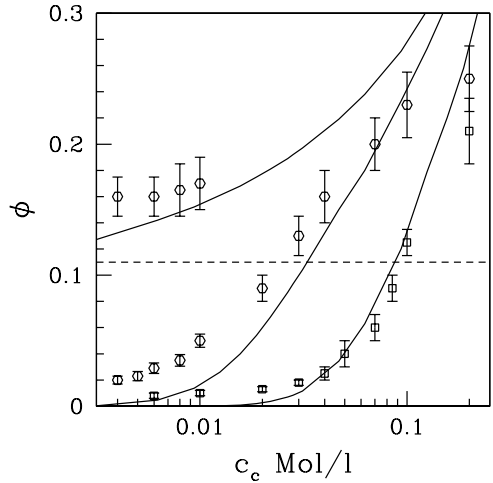,width=7cm}
\caption{Osmotic coefficient $\phi$ 
of B-DNA solutions as a function of density of DNA
phosphate ions $c_{c}$, for ionic strengths of 10 mM, 2mM and 0 mM (from
bottom to top). The dots are the experimental points obtained from Refs. 
{\protect\cite{BDNA1,BDNA2}}, while the solid lines correspond to the 
predictions for $\phi $ using our prescription in cylindrical geometry. 
The dashed line is the prediction of Oosawa-Manning condensation theory.}
\label{osm}
\end{figure}
\end{center}

For completeness, we compare in what follows 
our estimate for the osmotic pressure to the
experimental results on B-DNA solutions reported 
in \cite{BDNA1,BDNA2}. In this work, the
authors measured the osmotic coefficient 
$\phi =\Pi _{\hbox{\scriptsize osm}}/\Pi _{c}$, 
defined as the ratio between the osmotic
pressure $\Pi _{\hbox{\scriptsize osm}}$ to the pressure $\Pi _{c}$ of
releasable counter-ions having bare density $c_{c}$ ($\Pi _{c}=k_{B}Tc_{c}$)
against the concentration of  B-DNA, a rigid cylindrical polyelectrolyte.
A related PB cell analysis may be found in \cite{HansenPL} while 
a more thorough investigation has been performed in \cite{Ballauff}.

Within the WS model, B-DNA macro-ions are confined into cylindrical cells,
which radius $R_{WS}$ is related to the bare concentration of DNA
counter-ions as $c_{c}=(\ell _{DNA}\pi R_{WS}^{2})^{-1}$, with $\ell
_{DNA}=1.7\,$\AA\ the distance between charges along DNA. Note that as in
the previous section dealing with charged spheres, there is no
adjustable parameter in our description since the radius of the DNA and the
bare charge (only used to normalize the osmotic pressure to $\Pi _{c}$) are
known from independent measurements. 
In Fig. \ref{osm}, the corresponding results for the osmotic coefficient are
confronted against the experimental data of Refs. \cite{BDNA1,BDNA2} for various
ionic strengths, showing again a good quantitative agreement. As in Ref. 
\cite{Hansen}, we report the prediction of classical Oosawa-Manning
condensation theory (see e.g. \cite{Deserno,Ballauff}), for which 
the osmotic coefficient is constant [$\phi=\ell _{DNA}/(2\ell _{B})$] 
at complete variance with the experiments. In view of the results 
reported in Fig. \ref{fig:P_cyl}, the disagreement at small $c_c$
may be attributed to the (relative) failure of PB theory,
and not to a weakness of our prescription that should be judged with respect 
to non-linear PB.

%%%%%%%%%%%%%%%%%%%%%%%%%%%%%%%%%%%%%%%%%%%%%%%%%%%%%%%%%%%%%%%%%%%%%%%%%%%%%%
\section{Discussion - Validity of the approach}
\label{SectionDiscussion}

Our analysis was carried out at the level of Poisson-Boltzmann theory, 
which is mean-field like. More refined approaches such as the salt free 
Monte Carlo simulations of Groot \cite{Groot} for the cell model 
within the primitive model \cite{primitive}
have shown a non-monotonic behaviour of $\Zeff$ upon
increasing $Z$ for spheres: after the linear regime 
where $\Zeff \simeq Z$, $\Zeff$ reaches a maximum for a value
$Z_{\hbox{\scriptsize bare}}^*$ and then decreases.
When the radius $a$ of the charged spheres is much larger than Bjerrum
length $\lb$, this maximum is surrounded by a large plateau 
in excellent agreement with the PB saturation value $\Zsat$ \cite{Stevens,Groot}.
PB theory appears to become exact for $\lb/a \to 0$ \cite{rquefin}.
Given that, $Z_{\hbox{\scriptsize bare}}^*$ scales like $(a/\lb)^{2}$
and therefore becomes quickly large when the colloid size is 
increased \cite{Groot}, 
PB theory is successful in the colloidal limit of large $a$. 
We recall that this is precisely the limit where our predictions
for the effective charge at saturation $\Zsat$ are reliable
(the condition $a \gg \kappa^{-1}$ should {\it a priori} be satisfied
even if we have shown above that our predictions remain fairly
accurate down to $\kappa a$ of order 1).
More generally, the results of Groot \cite{Groot} show that the effective 
charge $\Zeff$ from Monte Carlo simulations within the primitive model 
for arbitrary $a/\lb$ are smaller than the quantity $\Zsat$ of PB theory.
This is a general feature that neglect of ionic correlation (as in PB)
leads to a underestimated screening of the macro-ion by the micro-ions,
and thus to overestimated effective charges (see e.g. \cite{Ballauff,Deserno2}
in cylindrical geometry). Our approach thus provides a useful
upper limit for a realistic $\Zeff$. 

A related comment in favor of the validity of PB picture with 
a saturation plateau for $\Zeff$ comes from the work of Cornu
and Jancovici \cite{Cornu}.
For the two dimensional two-component Coulomb gas bounded by a
hard wall of surface charge $\sigma $, 
these authors performed an exact calculation at the inverse
reduced temperature $e^{2}/(kT)=2$ showing that the effective surface charge
of the wall saturates to a plateau value when $\sigma $
diverges. 

Generally speaking, in an 1:1 electrolyte, PB theory seems to
be a reasonable approximation \cite{Hribar,Evilevitch}, 
all the better that the size of the macro-ion is
larger than $\lb$; the notion of charge renormalization then encaptures the main
effects of the non-linear PB theory, and is consistent with experimental data
on dilute bulk solution \cite{Palberg,Crocker2} (see also the experimental
work cited in section \ref{sectionExp}). For micro-ions of higher 
valences (di- or tri-valent), strong ionic correlations rule out 
PB-like approaches, as shown by recent computer simulations of the 
primitive model \cite{Allahyarov2,Linse,Messina}. As a consequence, 
the results presented here should {\it a priori}
not be used in the interpretation of experiments involving multivalent
salts or counter-ions. For a discussion concerning the effects of multivalent
counter-ions, we refer to the 
review by Bhuiyan, Vlachy and Outhwaite \cite{Bhuiyan}.

%%%%%%%%%%%%%%%%%%%%%%%%%%%%%%%%%%%%%%%%%%%%%%%%%%%%%%%%%%%%%%%%%%%%%%%%%%%%%%
\section{Conclusion}
\label{SectionConclusion}

The notion of effective charge is widely used 
in the fields of colloidal suspensions. It allows in practice to describe
the phase behavior of (highly) charged macro-ions staying at the level of
linearized Poisson-Boltzmann equations, where the macro-ions are supposed to
interact through Yukawa-like pair potentials. However, no general analytical
description of this renormalization process is available and the effective
charge is usually left as a free parameter, adjusted to fit experimental (or
numerical) data. 
Physically the charge renormalization process results from the strong
coupling of the micro-ions in the vicinity of the highly charged colloidal
particle. At the level of Poisson-Boltzmann theory, the effective charge 
saturates to a finite value in the limit where the bare charge becomes large.
We recall that omission of the non-linearities of PB theory
--correctly accounted for by the notion of effective charge--
may lead to unphysical results (see e.g. \cite{DesernoGrunberg,Grunberg}).

In the present paper, we have put forward a simple method to estimate
the effective charge of highly charged colloidal objects either
analytically, or through the resolution of a simple equation obtained within
linearized Poisson-Boltzmann approximation. This approach (mostly suited to
describe the colloidal limit $\kappa a \gg 1$) amounts to consider the
highly charged colloids as objects with constant electrostatic potential $%
\sim 4kT/e$, independently of shape and physico-chemical parameters (size,
added 1:1 electrolyte\ldots ). This result relies on the physical picture
that the electrostatic energy $e V_0$ of the strongly coupled micro-ions (%
\textit{i.e. } micro-ions in the vicinity of the highly charged macro-ion)
does balance their thermal (entropic) energy $k_BT$, resulting in a constant
effective surface potential for the ``dressed'' macro-ion. 
We have successfully tested this approach against 
\textit{a)} the geometry of the solid particle, \textit{b)} the confinement
(finite concentration situations), \textit{c)} the presence of added salt, 
\textit{d)} exact and approximate solutions of the
full non-linear PB equations \textit{e)} direct experimental measurements of
the effective charge found in the literature. From these different
checks, we conclude that our prescription appears to contain the key
ingredients involved in charge renormalization.

An important point is that the effective charge is not
constant and depends explicitly on the physical conditions of the
experiment, through ionic strength, density, etc \dots. The effect is quite
obvious in the small dilution limit, where we found that the (saturated)
effective
charge is an \textit{increasing} function of $\kappa$ (for $\kappa a>1$, 
which stems from the
reduction of the attraction between the counter-ions and the colloid. It 
pertains for finite concentration and the effective charge increases
with the ionic strength in the suspension. Addition of salt consequently
brings two antagonist effects on the effective Coulombic interaction between
macro-ions: the range of the interaction decreases due to screening, while
the amplitude increases due to the effective charge. The competition between
these two effects might be a key point in the understanding of these
systems. It appears therefore interesting to reconsider the phase stability
of macro-ion suspension in light of these results (see also 
\cite{Beresford} and more recently \cite{Gisler}).

Eventually it would be desirable to extend our approach to the case of finite
size colloidal particles, such as rods with finite length or 
disks \cite{Trizacbis}. Accordingly, edge effects should
show up at the level of our prescription and result in an effective charge
distribution along the macro-ion, due to the constant potential prescription
on the object. Work along these lines is in progress.

Acknowledgments: We would like to thank J.P. Hansen, H. L\"owen, 
H.H. von Gr\"unberg, M. Deserno, C. Holmand, J.F. Joanny for useful discussions.

%%%%%%%%%%%%%%%%%%%%%%%%%%%%%%%%%%%%%%%%%%%%%%%%%%%%%%%%%%%%%%%%%%%%%%%%%%%%%%%%
%%%%%%%%%%%%%%%%%%%%%%%%%%%%%%%%%%%%%%%%%%%%%%%%%%%%%%%%%%%%%%%%%%%%%%%%%%%%%%%%

%%%%%%%%%%%%%%%%%%%%%%%%%%%%%%%%%%%%%%%%%%%%%%%%%%%%%%%%%%%%%%%%%%%%%%%%%%%%%%%%%%%
%%%%%%%%%%%%%%%%%%%%%%%%%%%%%%%%%%%%%%%%%%%%%%%%%%%%%%%%%%%%%%%%%%%%%%%%%%%%%%%%%%%
\newpage

\begin{appendix}
\section{Analytical solution of PB equation for an isolated plane 
in an electrolyte}
\label{appendixA}

Here, we recall the analytical solution of PB equation for an isolated
plane of bare surface charge $\sigma e$ immersed in an electrolyte of bulk
ionic strength $I_{0}$. In this
case, the solution of equation (\ref{PBsalt}) reads \cite{Andelman}: 
\begin{equation}
\phi _{_{\hbox{\scriptsize PB}}}(z)=2\ln \frac{1+\gamma e^{-\kappa z}}{%
1-\gamma e^{-\kappa z}}
\end{equation}
where $\gamma =\sqrt{x^{2}+1}-x$, $\kappa ^{2}=8\pi \ell _{B}I_{0}$ and $%
x=\kappa \lambda _{GC}$, with the Gouy-Chapman length defined as
\begin{equation}
\lambda _{GC}=\frac{1}{2\pi l_{B}\sigma }.
\label{PotnonlinPlan}
\end{equation}

Far from the charged surface, say $z>2\kappa ^{-1}$, the solution of PB
equation reduces to 
\begin{equation}
\phi _{_{\hbox{\scriptsize PB}}}(z) \simeq \phi _{S}e^{-\kappa z}
\label{potapparentPlan}
\end{equation}
with $\phi _{S}=4\gamma $. The potential $\phi _{S}$ can be interpreted as
the \textit{apparent} reduced potential extrapolated at contact. In the
following we shall simply refer to $\phi _{S}$ as the \textit{apparent
potential}.

As expected, this asymptotic expression for the reduced potential 
$\phi _{_{\hbox{\scriptsize PB}}}$ in Eq. (\ref{potapparentPlan}) precisely matches
the solution $\phi _{_{\hbox{\scriptsize LPB}}}$ of the linearized PB (LPB)
equation: 
\begin{equation}
\nabla^2 \phi _{_{\hbox{\scriptsize LPB}}}=\kappa ^{2}
\phi _{_{\hbox{\scriptsize LPB}}}  
\label{LPBplan}
\end{equation}
but with the fixed charge boundary condition on the plane replaced by an
effective \textit{fixed surface potential} boundary condition $\phi _{_{%
\hbox{\scriptsize LPB}}}(z=0)=\phi _{S}=4\gamma $. The effective charge
density is then computed using Gauss theorem at the surface, yielding 
\begin{equation}
\sigma _{\hbox{\scriptsize eff}}=\frac{\gamma \kappa }{\pi l_{B}}.
\label{chargeeffplan}
\end{equation}
In doing so, we have replaced the initial non linear PB equation with fixed
charge boundary condition by the linear LPB equation with fixed surface
potential.

Now at fixed $\kappa $ (\textit{i.e.} constant ionic strength), we
progressively increase the bare surface charge $\sigma $. Accordingly $%
\kappa \lambda _{GC}\rightarrow 0$ and the parameter $\gamma $ goes to 1.
From Eq. (\ref{chargeeffplan}), we obtain that the effective
charge and the apparent potential $\phi _{S}$ have a simple behaviour
depending on the comparison of $\sigma $ with 
$\sigma _{\hbox{\scriptsize sat}}$ defined as 
\begin{equation}
\sigma _{\hbox{\scriptsize sat}}=\frac{\kappa }{\pi l_{B}}.
\label{eq:appsigmasat}
\end{equation}
Indeed 
\begin{eqnarray}
\sigma \ll \sigma _{\hbox{\scriptsize sat}}\qquad &&\sigma _{%
\hbox{\scriptsize eff}}\cong \sigma  \nonumber \\
&&\phi _{S}\cong 4\,\sigma /\sigma _{\hbox{\scriptsize sat}} \\
\sigma \gg \sigma _{\hbox{\scriptsize sat}}\qquad &&\sigma _{%
\hbox{\scriptsize eff}}\cong \sigma _{\hbox{\scriptsize sat}}  \nonumber \\
&&\phi _{S}\cong 4
\end{eqnarray}
The important point is that in the large bare charge limit, $\sigma \gg
\sigma _{\sat}$, the effective charge $\sigma_{\eff}$ saturates to
a value $\sigma_{\sat}$ independent of the bare one, $\sigma $. In this
limit, the apparent potential also saturate to a constant value, $\phi
_{S}=4 $.

%%%%%%%%%%%%%%%%%%%%%%%%%%%%%%%%%%%%%%%%%%%%%%%%%%%%%%%%%%%%%%%%%%%%%%%%%
\section{Analytical solution of PB equation for a confined plane 
without added salt} 
\label{appendixB}

An infinite plane (bare surface charge density $\sigma e$) 
is placed in the middle of a Wigner-Seitz 
slab of width $2h$. The origin of coordinates 
$x=0$ is chosen
at the location of the plane such that the volume available to the
counter-ions is $-h\leq x\leq h$. For symmetry reasons, it is enough
to solve the problem for $x>0$.
The electrostatic potential $\phi $ obeys PB equation  (\ref{PBnosalt}),
supplemented with Neumann boundary conditions: 
$\nabla \phi (h)=0$, corresponding to the electroneutrality condition; $%
\nabla \phi (0)=-2\pi \,\ell _{B}\sigma  $ imposing
the charge on the plane. Without loss of generality, we choose the origin
of potential such that $\phi (h)=0$; the analytical solution of  
PB equation then reads \cite{Andelman}
\begin{equation}
\phi _{_{\hbox{\scriptsize PB}}}(x)=-\log \left[ \cos ^{2}\left( \frac{%
\left( |x|-h\right) }{\sqrt{2}K_{_{\hbox{\scriptsize PB}}}^{-1}}\right)
\right] ,
\end{equation}
where $K_{_{\hbox{\scriptsize PB}}}(\sigma )$ is such that 
\begin{equation}
\frac{hK_{_{\hbox{\scriptsize PB}}}}{\sqrt{2}}\tan 
\left( \frac{hK_{_{\hbox{\scriptsize PB}}}}{\sqrt{2}}\right) =
\pi \ell _{B}\sigma h.
\label{eq:Kpbnosalt}
\end{equation}
The inverse screening length $K_{_{\hbox{\scriptsize PB}}}$ is related
to the density of counter-ions at the WS boundary $\rho^{-}(h)$
through the following expression, reminiscent of the standard definition
of the Debye length
\be
K_{_{\hbox{\scriptsize PB}}}^2 \,=\, 4 \pi \lb \, \rho^-(h).
\ee

Now we consider the corresponding LPB equation. More precisely, we linearize
Eq. (\ref{PBnosalt}) around $x=h$ (\textit{i.e.} the
edge of the slab). Since we have chosen 
$\phi_{_{\hbox{\scriptsize PB}}}(h)=0$, we impose 
$\phi_{_{\hbox{\scriptsize LPB}}} $ to vanish at $x=h$. 
The resulting equation reads 
\begin{equation}
\nabla ^{2}\phi =K_{_{\hbox{\scriptsize LPB}}}^{2}\left( \phi +1\right)
\label{eq:lpb}
\end{equation}
where we have introduced $K_{_{\hbox{\scriptsize LPB}}}$, an ``apparent''
local Debye screening factor for the linearized PB equation. 
As for the previous PB equation in the no salt case, 
$K_{_{\hbox{\scriptsize LPB}}}$ is not known \textit{a priori} 
but results from the
electroneutrality condition. Indeed, solving Eq. (\ref{eq:lpb}) with the
appropriate boundary conditions [$\nabla \phi (h)=0$, $\nabla \phi (0)=4\pi
\,\ell _{B}\left( \sigma /2\right) $] yields 
\begin{equation}
\philpb(x)=\cosh \left[ K_{_{\hbox{\scriptsize LPB}}}\left(
x-h\right) \right] -1,  \label{pot_LPB}
\end{equation}
with $K_{_{\hbox{\scriptsize LPB}}}(\sigma )$ such that 
\begin{equation}
hK_{_{\hbox{\scriptsize LPB}}}\sinh \left[ K_{_{\hbox{\scriptsize LPB}}}
h\right] =2\pi \ell _{B}\sigma h  \label{eq:Klpbnosalt}
\end{equation}
Note that comparing Eqs. (\ref{eq:Kpbnosalt}) and (\ref{eq:Klpbnosalt}), we
see that $K_{_{\hbox{\scriptsize LPB}}}(\sigma)\neq 
K_{_{\hbox{\scriptsize PB}}}(\sigma)$. It is however crucial to remember
that the LPB solution should not be used with the bare charge $\sigma$ 
to describe the correct behaviour of $\phipb$ in the vicinity of $x=h$.

Next, we implement the procedure proposed by Alexander to find the effective
charge in confined situations \cite{Alexander}. The
effective charge density is accordingly 
the value of $\sigma $ in the linearized PB
equation such that $\phi _{_{\hbox{\scriptsize PB}}}(x)$ and $\phi _{_{%
\hbox{\scriptsize LPB}}}(x)$ match up to the second derivative at $\Sigma $,
the boundary of the WS cell \cite{Alexander}.
This condition is equivalent to set 
\begin{equation}
K_{_{\hbox{\scriptsize LPB}}}\left( \sigma_{\eff}\right) =K_{_{%
\hbox{\scriptsize PB}}}(\sigma ).  
\label{eq:K=K}
\end{equation}
Note that in general, whenever the solution of the non-linear PB problem is
known, the effective charge $\sigma_{\eff}$ can be directly estimated with
Eq. (\ref{eq:K=K}) (this is of course quite academic to obtain in this case
an effective charge since the full solution for the potential is known; the
notion of effective charge is mostly useful in geometries where no
analytical solution of the PB equation is known). Note also that whenever
Eq. (\ref{eq:K=K}) is verified, the third, fourth and fith derivative 
of the linear and
non-linear solutions also match at $\Sigma $.

One deduces eventually the ``exact'' effective charge from Eq. (\ref{eq:K=K}%
) for the plane case (by ``exact'' we mean that the effective charge is
obtained from the analytical solution of PB equation, in contrast to our
prescription): 
\begin{equation}
\sigma_{\eff}=\frac{K_{_{\hbox{\scriptsize PB}}}(\sigma )\sinh \left[ K_{_{%
\hbox{\scriptsize PB}}}(\sigma )h\right] }{2\pi \ell _{B}}.
\end{equation}
The apparent potential $\phi _{S}$ is also obtained as 
\begin{equation}
\phi _{S}=\philpb(0)=\cosh \left[ K_{_{\hbox{\scriptsize PB}}}(\sigma
)h\right] -1.
\end{equation}
From Eq. (\ref{eq:Kpbnosalt}), we define a critical value for the charge
density 
\begin{equation}
\sigma _{c}=\frac{1}{\pi l_{B}h}
\end{equation}
and we find the asymptotic behaviors 
\begin{eqnarray}
\sigma  &\ll &\sigma _{c}\qquad \left\{ 
\begin{array}{l}
hK_{_{\hbox{\scriptsize PB}}}\cong (2\sigma /\sigma _{c})^{1/2} \\ 
\sigma_{\eff}\cong \sigma  \\ 
\phi _{S}\cong 2\,\sigma /\sigma _{c}
\end{array}
\right.  \\
\sigma  &\gg &\sigma _{_{c}}\qquad \left\{ 
\begin{array}{l}
hK_{_{\hbox{\scriptsize PB}}}\cong \pi /\sqrt{2} \\ 
\sigma_{\eff} \cong \sigma_{\sat} = \frac{\pi }{2\sqrt{2}}\sinh \left[ 
\frac{\pi }{\sqrt{2}}\right] \sigma _{c} \cong 5.06 \sigma _{c} \\ 
\phi _{S}\cong \cosh \left[ \frac{\pi }{\sqrt{2}}\right] -1\cong 3.66
\end{array}\right.   
\label{sigma_sat}
\end{eqnarray}

As in the infinite dilution limit, one obtains that the
effective charge $\sigma_{\eff}$ coincides with the bare one $\sigma $ for
small $\sigma $, and saturates to a finite value 
when $\sigma \to \infty $. However both $\sigma _{c}$ and the saturation value
for the effective charge at finite concentration differ from the
$\sigma_{\sat}$ of infinite dilution [Eq. (\ref{eq:appsigmasat})].
We also note that strictly speaking, 
the limits of infinite dilution and of vanishing added
salt do not commute: if the limit of vanishing salt is taken first
(situation investigated in this appendix), before $h\to \infty$,
we obtain $\phi_S^{\sat} = \cosh \left[ \pi /\sqrt{2}\right] -1$,
whereas reverting the order corresponds to the planar situation
of Appendix \ref{appendixA} with $\kappa \to 0$, and there,
we have $\phi_S^{\sat} = 4$.  In both cases however, the effective charge 
at saturation vanishes.

\end{appendix}

%\end{multicols}


\begin{references}
\bibitem{rem} In this respect, it is important to
realize that for an infinitely long rod in the Manning limit,
the effective line charge $\laeff$ is smaller than the equivalent charge
$\lambda_{\hbox{\scriptsize equiv}}$. For large bare charges, we have 
in particular $\lasat < 1/\lb$, see the discussion in section \ref{sectionZeff1}.
\bibitem{primitive}
which neglects the discrete nature of the solvent,
but takes into account the micro-ions explicitly as hard bodies.
\bibitem{rquefin}
with the mathematical restriction that $Z \lb/a$ remains finite.
\end{references}
\end{document}